\documentclass[%
 reprint,
superscriptaddress,
nofootinbib,
 amsmath,amssymb,
 aps,
 prd,
floatfix,
]{revtex4-2}
\usepackage{graphicx}

\usepackage{xcolor}

\usepackage{hyperref}
\definecolor{plasmablue}{rgb}{0.050383, 0.029803, 0.527975}
\hypersetup{
  colorlinks=true,
  urlcolor=plasmablue,
  citecolor=plasmablue,
  }

\newcommand{\ie}{i.\,e.\ }
\newcommand{\eg}{e.\,g.\ }

\newcommand{\dd}{\textnormal{d}}

\newcommand{\asz}{\ensuremath{{\zeta_0}}}
\newcommand{\bsz}{\ensuremath{{\zeta_M}}}
\newcommand{\csz}{\ensuremath{{\zeta_z}}}
\newcommand{\sigmalnzeta}{\ensuremath{{\sigma_{\ln\zeta}}}}
\newcommand{\alambda}{\ensuremath{{\lambda_0}}}
\newcommand{\blambda}{\ensuremath{{\lambda_M}}}
\newcommand{\clambda}{\ensuremath{{\lambda_z}}}
\newcommand{\sigmalnlambda}{\ensuremath{{\sigma_{\ln\lambda}}}}

\newcommand{\bWL}{\ensuremath{{\ln M_{\mathrm{WL}_0}}}}
\newcommand{\bWLM}{\ensuremath{{M_{\mathrm{WL}_M}}}}
\newcommand{\sWL}{\ensuremath{{\ln\sigma^2_{\ln\mathrm{WL}_0}}}}
\newcommand{\sWLM}{\ensuremath{\sigma^2_{\ln\mathrm{WL}_M}}}

\begin{document}

\title{Constraining \texorpdfstring{$f(R)$}{f(R)} gravity using future galaxy cluster abundance and weak-lensing mass calibration datasets}

\author{Sophie M.\ L.\ Vogt}
\email{s.vogt@physik.lmu.de}
\affiliation{
University Observatory, Faculty of Physics, Ludwig-Maximilians-Universit\"at, Scheinerstr.\ 1, 81679 Munich, Germany
}
\affiliation{
Excellence Cluster ORIGINS, Boltzmannstr.\ 2, 85748 Garching, Germany
}
\affiliation{Max Planck Institute for Astrophysics, Karl-Schwarzschild-Str.\ 1, 85748 Garching, Germany}
\author{Sebastian Bocquet}
\affiliation{
University Observatory, Faculty of Physics, Ludwig-Maximilians-Universit\"at, Scheinerstr.\ 1, 81679 Munich, Germany
}
\author{Christopher T.\ Davies}
\affiliation{
University Observatory, Faculty of Physics, Ludwig-Maximilians-Universit\"at, Scheinerstr.\ 1, 81679 Munich, Germany
}
\author{Joseph J.\ Mohr}
\affiliation{
University Observatory, Faculty of Physics, Ludwig-Maximilians-Universit\"at, Scheinerstr.\ 1, 81679 Munich, Germany
}
\affiliation{Max Planck Institute for extraterrestrial Physics, Giessenbach-Str.\ 2, 85748 Garching, Germany}
\author{Fabian Schmidt}
\affiliation{Max Planck Institute for Astrophysics, Karl-Schwarzschild-Str.\ 1, 85748 Garching, Germany}
\affiliation{
Excellence Cluster ORIGINS, Boltzmannstr.\ 2, 85748 Garching, Germany
}

\begin{abstract}
We present forecasts for constraints on the Hu \& Sawicki $f(R)$ modified gravity model using realistic 
mock data representative of future cluster and weak lensing surveys. We create mock thermal Sunyaev-Zel'dovich effect selected cluster samples for SPT-3G and CMB-S4 and the corresponding weak gravitational lensing data from next-generation weak-lensing (ngWL) surveys like Euclid and Rubin. We employ a state-of-the-art Bayesian likelihood approach that includes all observational effects and systematic uncertainties to obtain constraints on the $f(R)$ gravity parameter $\log_{10}|f_{R0}|$. In this analysis we vary the cosmological parameters $[\Omega_{\rm m}, \Omega_\nu h^2, h, A_s, n_s, \log_{10}|f_{R0}|]$, which allows us to account for possible degeneracies between cosmological parameters and $f(R)$ modified gravity. The analysis accounts for $f(R)$ gravity via its effect on the halo mass function which is enhanced on cluster mass scales compared to the expectations within general relativity (GR). Assuming a fiducial GR model, the upcoming cluster dataset SPT-3G$\times$ngWL is expected to obtain an upper limit of $\log_{10}|f_{R0}| < -5.95$ at $95\,\%$ credibility, which significantly improves upon the current best bounds. The CMB-S4$\times$ngWL dataset is expected to improve this even further to $\log_{10}|f_{R0}| < -6.23$. 
Furthermore, $f(R)$ gravity models with $\log_{10}|f_{R0}| \geq -6$, which have larger numbers of clusters, would be distinguishable from GR with both datasets. We also report degeneracies between $\log_{10}|f_{R0}|$ and $\Omega_{\mathrm{m}}$ as well as $\sigma_8$ for $\log_{10}|f_{R0}| > -6$ and $\log_{10}|f_{R0}| > -5$ respectively. Our forecasts indicate that future cluster abundance studies of $f(R)$ gravity will enable substantially improved constraints that are competitive with other cosmological probes.
\end{abstract}

\maketitle


\section{\label{sec:intro}Introduction}
The cause of the late-time accelerated expansion of the universe is one of the most pertinent and challenging questions in cosmology \cite{Perlmutter1999,Riess1998}. The phenomenon can be explained within the framework of general relativity (GR) if a fluid with negative pressure is introduced, or alternatively if a cosmological constant $\Lambda$ is added to the Einstein-Hilbert action. The latter scenario is known as the $\Lambda$ cold dark matter model, $\Lambda$CDM. An alternative approach to explain cosmic acceleration is through the modification of the Einstein-Hilbert action, in a manner which seeks to avoid the addition of a cosmological constant, referred to as modified gravity. 
A key feature of modified gravity is that 
the clustering of matter is in general different from that of GR \citep[see, e.g.,][for reviews]{JoyceEtal16,Koyama18,Baker19}. 
This means that studies of the growth of structure can be a useful test of GR on cosmic scales, which may then shed light on the underlying cause of the late-time acceleration. 

One of the most popular modified gravity models alters the Einstein-Hilbert action 
using a non-linear function $f(R)$ of the scalar curvature $R$ \cite{Buchdahl1970}. This leads to an additional, gravitational-strength, fifth force. This fifth force affects 
structure formation, introducing a scale-dependence and enhancing the structure growth on galaxy cluster scales. 
A further feature of $f(R)$ gravity models is the chameleon screening mechanism, which suppresses the modification in high-density regions \cite{Khoury04}, ensuring that $f(R)$ remains consistent with tests of GR in the Solar system \cite{Burrage18}. 
In this work we use the widely studied $f(R)$ model of Hu \& Sawicki \cite{Hu07} with $n= 1$ for which self-consistent $N$-body simulations \cite{Oyaizu08,Li11,Cataneo16,Arnold21} as well as semi-analytical models for the halo mass function \cite{Lombriser13,Lombriser14} exist.

The abundance of massive galaxy clusters and their evolution over cosmic time has long been recognized as a powerful probe of the cosmic acceleration \cite{Wang98,Haiman01}. 
Cluster surveys have been used to constrain $\Lambda$CDM \cite{Vikhlinin09,Benson13,Bocquet15,Dehaan16,Bocquet19,Abbott20,Bocquet24} as well as modified gravity models \cite{Schmidt09b,Lombriser10,Cataneo14,Peirone17,Hagstotz19,Artis24}. 
The abundance of collapsed haloes is described by the halo mass function (HMF), which, through the scale-dependent structure growth, depends on the strength of the $f(R)$ gravity model, 
where the strength of the model is encoded in the single parameter $f_{R0}$. 
This, together with the nonlinear screening mechanism, results in a mass-dependent enhancement of the HMF. We incorporate this via a prefactor which depends on the critical overdensity of halo collapse in $f(R)$ gravity \cite{Cataneo14}. 
The critical overdensity in $f(R)$ gravity can be calculated via a semi-analytical model \citep{Li12a,Lombriser13}. In this approach, the critical overdensity is computed for each halo mass and redshift by solving a system of coupled differential equations. This makes the calculation computationally expensive, and so we construct an emulator to make the computation more efficient.

In practice, the mass of a galaxy cluster, and hence the HMF is not measured directly. Instead, observable signatures are used to infer the cluster mass.
These observables are measured and then mapped to the halo mass through observable--mass relations, which are generally of power-law form in mass and redshift and include a model of the scatter in the observable at a fixed halo mass and redshift. 
To calibrate these relations, a robust mass measurement is needed.
In this analysis, we employ weak gravitational lensing data (WL) for the mass calibration of the cluster ensemble. 
The advantage of using the WL signal to calibrate the halo masses is that the WL data provide mass constraints with well-characterized and controllable biases and associated uncertainties on those biases. Moreover, no assumption of hydrostatic or virial equilibrium is required.

One way to detect massive galaxy clusters is through the thermal Sunyaev-Zel’dovich effect (tSZE) \cite{Sunyaev1972}. This phenomenon arises from the upscattering of the cosmic microwave background (CMB) photons by energetic electrons within the intracluster medium (ICM). 
The resulting spectral distortion of the CMB is redshift independent, whereas the observable signature is approximately independent of redshift. This contributes to the fact that tSZE surveys provide a clean probe to study the growth of structure up to the highest redshifts where massive clusters exist (\ie, $z \simeq 2$).

Ongoing and planned tSZE surveys such as those from SPT-3G \cite{Benson14}, conducted with the South Pole Telescope (SPT) \cite{Carlstrom11}, Simons Observatory \cite{Ade19}, or 
CMB-S4 \cite{Abazajian19} will detect thousands to tens of thousands of galaxy clusters \cite{Raghunathan22}. 
Datasets from next-generation galaxy weak-lensing (hereafter ngWL) surveys as from the ongoing Euclid mission \cite{Laureijs11,Scaramella22} or the Vera C.\ Rubin Observatory \cite{Ivezic08,Mandelbaum18} will allow for improved and more robust measurements of cluster halo masses. Therefore, combining future cluster tSZE survey data with ngWL data will yield powerful probes 
of cosmology and modified gravity,
greatly improving upon the existing constraints from cluster abundance analyses. 

Modified gravity models such as $f(R)$ in general not only modify the halo mass function, but also the observable--mass relation \cite{Schmidt2010} and the halo profiles \cite{Ruan23}. This results in slightly different values of the weak-lensing mass to halo mass relation parameters, relative to those in GR.  For the models considered in this work, these effects are small, and we therefore assume that they are accounted for within the systematic uncertainty budget assigned to the weak-lensing mass calibration.

In this work, we make realistic forecasts for the constraining power of weak-lensing informed galaxy cluster abundance studies on $f(R)$ gravity. We use tSZE surveys from SPT-3G and CMB-S4 combined with cluster mass calibration constraints from ngWL datasets similar to those expected from Euclid and Rubin.
For this, we create mock cluster and WL data for different values of $\log_{10}|f_{R0}|$ as well as for a GR cosmology. We analyze the mock datasets with a cluster-by-cluster likelihood approach in two steps. First, we use the ngWL dataset for the full cluster sample to constrain the observable--mass relation parameters, and then 
we adopt the posteriors from that first calculation as priors and employ the cluster abundance likelihood to obtain constraints on the $f(R)$ gravity parameter $\log_{10}|f_{R0}|$ and other cosmological parameters of interest.

The analysis presented in this work employs the state-of-the-art framework developed for the cosmological analyses of galaxy clusters selected in the SPT-SZ and SPTpol surveys, with weak-lensing mass calibration using data from the Dark Energy Survey data and the Hubble Space Telescope (HST) \citep{Bocquet23}. This analysis framework and dataset was used to obtain competitive cosmological constraints on $\Lambda$CDM and $w$CDM cosmologies \cite{Bocquet24}. 
Constraining $f(R)$ gravity using SPT clusters with DES and HST weak-lensing is in progress and will presented in a future work.

The paper is organized as follows. Section~\ref{sec:MG} presents the $f(R)$ modified gravity model, briefly describing the Hu \& Sawicki model and discusses the semi-analytical spherical collapse model used to calculate the critical overdensity and subsequently quantify how the HMF changes with respect to the GR version. Moreover, we present emulators which are used to speed up the calculation of the critical overdensity and its derivative with respect to $\mathrm{ln}M$. We describe in Section~\ref{sec:surveys} the different cosmological surveys we use for the forecasts.
The observable--mass relations used in our analysis are summarized in Section~\ref{sec:obs-mass-rel}.
The generation of the mocks is described in Section~\ref{sec:mocks}.
We present the likelihood and the analysis method in Section~\ref{sec:likel_analysis}.
The results of our analysis are presented in Section~\ref{sec:results}. Finally, Section~\ref{sec:discussion} gives a brief summary of our work.

Throughout this paper $\mathcal{U}(a, b)$ denotes a uniform distribution between limits $a$ and $b$, and $\mathcal{N}(\mu, \sigma)$ is a Gaussian distribution with mean $\mu$ and standard deviation $\sigma$. 
In this analysis we adopt the halo mass definition $M_{200\mathrm{c}}$, which is the mass within the cluster region where the enclosed mean density is 200 times the critical density.

\section{\label{sec:MG}\texorpdfstring{$f(R)$}{f(R)} Modified gravity}
In $f(R)$ gravity models the Einstein-Hilbert action of general relativity (GR) is modified to include an arbitrary function $f(R)$ of the scalar curvature $R$ \cite{Buchdahl1970}
    \begin{equation}
        \label{eq:fR_EH_action}
        S = \int \dd^4 x \sqrt{-g} \left [ \frac{R + f(R)}{16\pi G} + \mathcal{L}_m \right ] \, .
    \end{equation}
Here $g$ is the determinant of the GR metric tensor $g_{\mu \nu}$, $G$ the gravitational constant and $\mathcal{L}_m$ is the matter Lagrangian. Note, that we use natural units $c = \hbar = 1$. The field equation for $f(R)$ gravity can be obtained by varying the action with respect to the metric tensor
    \begin{equation}
        \label{eq:fR_field_eq}
        \begin{split}
            G_{\mu \nu} + f_R R_{\mu \nu} -& \left (\frac{f}{2} - \square f_R \right ) g_{\mu \nu} \\
            & \hspace{1.5cm} - \nabla_\mu \nabla_\nu f_R  = 8 \pi G T_{\mu \nu} \, ,
        \end{split}
    \end{equation}
where $G_{\mu \nu}$ denotes the Einstein tensor, $R_{\mu \nu}$ represents the Ricci tensor, $T_{\mu \nu}$ is the energy-momentum tensor and $f_R = \dd f(R) / \dd R$ which behaves as an additional scalar degree of freedom and is named the scalaron field.

Under the quasistatic and weak-field approximation, the trace of the field equation gives the equation of motion for $f_R$
    \begin{equation}
        \label{eq:EoM_fR}
        \nabla^2 \delta f_R = \frac{1}{3} (\delta R - 8 \pi G \delta \rho) \, ,
    \end{equation}
with $\delta \rho = \rho - \bar \rho$. The modified Poisson equation is obtained from the time-time component of Eq.~\eqref{eq:fR_field_eq} 
    \begin{equation}
        \label{eq:poisson_eq}
        \nabla^2 \Phi = \frac{16 \pi G}{3} \delta \rho - \frac{1}{6} \delta R \,,
    \end{equation}
with $\delta R = R - \bar R$ and the Newtonian potential $\Phi$ is defined via $2\Phi = \delta g_{00} / g_{00}$. 
Combining these two equations gives for the modified Poisson equation
    \begin{equation}
        \label{eq:poisson_eq_2}
        \nabla^2 \Phi = 4 \pi G \delta \rho - \frac{1}{2} \nabla^2 \delta f_R \,.
    \end{equation}
So, in $f(R)$ gravity the Poisson equation includes an extra term directly proportional to $\nabla^2 \delta f_R$. 

To be effective at late times and large scales in cosmology, $f(R)$ has to be a decreasing function of $R$, so that $f_R < 0$.
Depending on the field value $f_R$ we can distinguish two different regimes: the large-field regime where $|f_R| \gg |\Phi|$, which corresponds to low curvature, and the small-field regime, \ie $|f_R| \ll |\Phi|$ and thus is related to high curvature \cite[see][for a detailed explanation]{Cataneo16,Hu07}.
In the first case, $\delta R \ll 8\pi G \delta \rho$ and the Poisson Eq.~\eqref{eq:poisson_eq} corresponds to an enhancement of $\Phi$, and hence gravitational forces, by a factor of $4/3$. On the other hand, in the small field regime, \ie large curvature, we have $\delta R \approx 8\pi G \delta \rho$ and thus Eq.~\eqref{eq:poisson_eq} reduces to the GR Poisson equation. Hence, structure growth becomes scale-dependent. The recovery of the GR Poisson equation in the high curvature regimes in $f(R)$ gravity is induced by the so-called chameleon screening mechanism \cite{Khoury04} which makes $f(R)$ gravity models consistent with solar system tests \cite{Burrage18}.

In this paper we use the Hu \& Sawicki form for the function $f(R)$ which is given by \cite{Hu07}
    \begin{equation}
        \label{eq:fR_Hu07}
        f(R) = - m^2 \frac{c_1\left( \frac{R}{m^2} \right)^n}{c_2\left( \frac{R}{m^2} \right)^n + 1} \, ,
    \end{equation}
with $m^2 = \Omega_\mathrm{m} H^2_0$, the free parameters $n$, $c_1$, $c_2$ and $H_0$ the Hubble constant. When $c_2^{1/n}R/m^2 \gg 1$, the above equation is approximately
    \begin{equation}
        \label{eq:fR_approx_high_curv}
        f(R) \approx - m^2 \frac{c_1}{c_2} - \frac{f_{R0} R_0^{n+1}}{nR^n} \, ,
    \end{equation}
where $R_0$ is the present background curvature and $f_{R0} := f_R(R_0)$ which is a parameter that quantifies the strength of the $f(R)$ gravity model.
Note that $f_{R0} < 0$. This approximation is correct up to order $\sim (f_{R0})^2$; since current constraints are already at the level of $|f_{R0}| \lesssim 10^{-4}$ or better, Eq.~\eqref{eq:fR_approx_high_curv} is entirely sufficient. Since we do not have a strong theory prior on a particular scale of $f_{R0}$, we choose to work with $\log_{10}|f_{R0}|$ for numerical convenience. We will also impose a uniform prior on $\log_{10}|f_{R0}|$ in our analysis. 

Since we recover GR in the limit $|f_{R0}| \to 0$, we obtain from Eq.~\eqref{eq:fR_EH_action}
    \begin{equation}
        \label{eq:condi_params_fR}
        \frac{c_1}{c_2} = 6 \frac{\Omega_\Lambda }{\Omega_{\mathrm{m}} } \, .
    \end{equation}
In this paper, we further adopt $n=1$, as this is by far the most widely studied $f(R)$ scenario; see Ref.~\cite{Ferraro2011} for an approach to approximately rescale constraints from $n=1$ to other values of $n$.

\subsection{\label{subsec:collapse} Spherical collapse in \texorpdfstring{$f(R)$}{f(R)} gravity}
As described in the previous section, structure growth is scale-dependent in $f(R)$ gravity, and consequently the formation of dark matter halos is too. 
For the formation of clusters we use a spherical collapse model developed and described in detail elsewhere~\cite{Li12a,Lombriser13,Cataneo16}. 

We approximate a dark matter halo as a spherical top-hat overdensity, with a radius $R_{\mathrm{TH}}$ and constant density inside $\rho_{\mathrm{in}}$ and outside $\rho_{\mathrm{out}}$. The mass and environment dependence of the spherical collapse model \cite{Li12a} with the chameleon screening effect is implemented through the thin-shell 
approximation \cite{Khoury04}. 

The physical radius of the spherical top-hat halo is defined as $\xi(a)$, where $\xi(a_{\mathrm{i}}) = a_{\mathrm{i}} R_{\mathrm{TH}}$ at initial scale $a_{\mathrm{i}}$. Due to the nonlinear evolution of the overdensity, the radius $\xi(a)$ deviates from the linear relation for larger scale factor $a$ and we define the variable $y$ to be the deviation from this linear relation, $y(a)= \xi(a)/aR_{\mathrm{TH}}$. Note that due to mass conservation, we have $\rho/\Bar{\rho}= y^{-3}$. The equation of motion for the spherical shell is given by \cite{Li12a,Schmidt09,Lombriser13}
    \begin{equation}
        \label{eq:EoM_sph_shell}
        \frac{\Ddot{\xi}}{\xi} = - \frac{4\pi G}{3}(\Bar{\rho}_m - 2\Bar{\rho}_\lambda) - \frac{4\pi G}{3} (1+F)\delta \rho_{\mathrm{m}} \,, 
    \end{equation}
where dots denote derivatives with respect to time. Here $F$ is the extra force from modified gravity and is given by the thin-shell approximation \cite{Khoury04,Lombriser13,Cataneo16}
\begin{small}
    \begin{equation}
        \label{eq:thin_shell_force}
        F \left ( \frac{\Delta \xi}{\xi} \right) = \frac{1}{3} \mathrm{min}\left( 3\frac{\Delta \xi}{\xi} - 3 \left ( \frac{\Delta \xi}{\xi} \right)^2 + \left ( \frac{\Delta \xi}{\xi} \right)^3, 1 \right) \, .
    \end{equation}
\end{small}
Here $\frac{\Delta \xi}{\xi}$ is the thickness of the thin shell 
and can be expressed as~\cite[see][]{Khoury04,Lombriser13}

    \begin{align}
        \label{eq:thin_shell}
        \frac{\Delta \xi}{\xi} \approx & \frac{|f_{R0}| a^{3n+4}}{\Omega_{\mathrm{m}}(H_0R_{\mathrm{TH}})^2} y_{\mathrm{h}} \Biggl[  \biggl( \frac{1+4 \frac{\Omega_\Lambda}{\Omega_{\mathrm{m}}}}{y_{\mathrm{env}}^{-3} + 4\frac{\Omega_\Lambda}{\Omega_{\mathrm{m}}}a^3  } \biggr)^{n+1} \nonumber \\
        & \hspace{2.1cm} -  \biggl( \frac{1+4 \frac{\Omega_\Lambda}{\Omega_{\mathrm{m}}}}{y_{\mathrm{h}}^{-3} + 4\frac{\Omega_\Lambda}{\Omega_{\mathrm{m}}}a^3  } \biggr)^{n+1} \Biggr] \, ,
    \end{align}
where $y_{\mathrm{h}}^{-3}$,  $y_{\mathrm{env}}^{-3}$ track the inner and outer overdensities respectively and $n=1$ due to our choice of the Hu~\&~Sawicki model. The evolution for the inner overdensity $y_{\mathrm{h}}^{-3}$ can be expressed as \cite[see][]{Lombriser13}:
    \begin{equation}
        \label{eq:EoM_inner}
        \begin{split}
        y_{\mathrm{h}}^{''} &+ \left( 2 - \frac{3}{2} \Omega_{\mathrm{m}}(a) \right) y_{\mathrm{h}}^{'} \\ 
        & \hspace{1.8cm} + \frac{1}{2}  \Omega_{\mathrm{m}}(a) (1 + F)(y_{\mathrm{h}}^{-3} - 1)y_{\mathrm{h}} = 0 \, ,
        \end{split}
    \end{equation}
with primes denoting derivatives with respect to $\mathrm{ln}a$. For the outer overdensity, we assume that it follows a GR evolution, thus $F \to 0$ and one obtains \cite{Lombriser13}
    \begin{equation}
        \label{eq:EoM_outer}
        \begin{split}
        y_{\mathrm{env}}^{''} &+ \left( 2 - \frac{3}{2} \Omega_{\mathrm{m}}(a) \right) y_{\mathrm{env}}^{'} \\ 
        & \hspace{1.8cm} + \frac{1}{2}  \Omega_{\mathrm{m}}(a) (y_{\mathrm{env}}^{-3} - 1)y_{\mathrm{env}} = 0 \, .
        \end{split}
    \end{equation}
Because the equation of motion for the inner region of the top-hat region depends on $F$ and thus by Eq.~\eqref{eq:thin_shell} on the outer region, Eqs.~\eqref{eq:EoM_inner} and \eqref{eq:EoM_outer} are a system of coupled differential equations. The initial conditions at $a_{\mathrm{i}} \ll 1$ are in the matter-dominated regime and given by
    \begin{align}
        \label{eq:initial_condi}
        y_{\mathrm{h/env, i}} &= 1 - \frac{\delta_{\mathrm{h/env, i}}}{3} \, , \\
        y^{'}_{\mathrm{h/env, i}} &= - \frac{\delta_{\mathrm{h/env, i}}}{3}\, .
    \end{align}
If we want to find the critical overdensity $\delta_{\mathrm{crit}}$ which causes spherical collapse at scale factor $a_{\mathrm{c}}$, we have to choose the initial conditions $y_{\mathrm{h, i}}$ and $y^{'}_{\mathrm{h, i}}$ such that the solution of Eq.~\eqref{eq:EoM_inner} gives $y_{\mathrm{h}}(a_{\mathrm{c}}) = 0$ with the requirement $y_{\mathrm{h}}(a) > 0$ for $a < a_{\mathrm{c}}$. The critical overdensity is then defined as the linearly extrapolated value of $\delta_{\mathrm{h/env, i}}$, \ie $\delta_{\mathrm{crit}} = D(a)/D(a_{\mathrm{i}}) \delta_{\mathrm{h/env, i}}$ with the linear growth factor $D(a)$. 

The initial overdensity $\delta_{\mathrm{env, i}}$ of the outer region is set by the peak of the probability distribution of an Eulerian environmental density $\delta_{\mathrm{env}}$ with an Eulerian (physical) radius $\xi = 5\,h^{-1}\,\mathrm{Mpc}$~\cite[for more details see][]{Lombriser13,Li12b} and is derived from a physical model to be \cite{Li12b} 
    \begin{align}
        \label{eq:eulerian_env}
        P_{\xi} (\delta_{\mathrm{env}})= &\frac{\beta^{\omega/2}}{\sqrt{2\pi}} \left[ 1+ (\omega - 1) \frac{\delta_{\mathrm{env}}}{\delta_{\mathrm{crit, \mathrm{GR}}}} \right] \left( 1 -   \frac{\delta_{\mathrm{env}}}{\delta_{\mathrm{crit,\Lambda}}} \right)^{\omega/2 -1} \nonumber \\
        &\times \exp \left( - \frac{\beta^{\omega}\delta_{\mathrm{env}} }{2(1-\delta_{\mathrm{env}} / \delta_{\mathrm{crit, \mathrm{GR}}})^{\omega}} \right) \, ,
    \end{align}
with $\beta = (\xi/8)^{3/\delta_{\mathrm{crit, \mathrm{GR}}}} (\sigma_8^{\rm GR})^{-2/\omega}$, $\delta_{\mathrm{crit, \mathrm{GR}}}$ the linear extrapolated GR overdensity and $\omega = \delta_{\mathrm{crit, \mathrm{GR}}} \gamma$ with 
    \begin{equation}
        \label{eq:gamma}
        \gamma = \frac{n_{\mathrm{s}} +3}{3}\, .
    \end{equation}

\begin{figure}
        \centering
        \includegraphics[width=\linewidth]{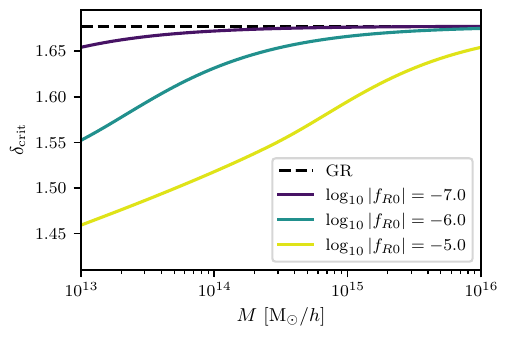}
        \vskip-0.25cm
        \caption{The critical overdensity $\delta_{\mathrm{crit}}$ for spherical collapse in $f(R)$ gravity Eq.~\eqref{eq:EoM_inner} for different values of $\log_{10}|f_{R0}|$ at collapse redshift $z_{\mathrm{c}} = 0$ in colored solid lines. The dashed black line represents $\delta_{\mathrm{crit}}$ in a corresponding GR cosmology Eq.~\eqref{eq:EoM_outer}. }
        \label{fig:delta_crit_semi}
    \end{figure}

In $f(R)$ gravity the critical overdensity for halo collapse $\delta_{\mathrm{crit}}$ is a function of the mass of the spherical overdensity $M = 4/3 \pi \Bar{\rho}_{\mathrm{m}}R^3_{\mathrm{RT}}$ via Eq.~\eqref{eq:thin_shell_force}, \eqref{eq:thin_shell} and the local overdensity of the environment $\delta_{\mathrm{env}}$ is given by Eq.~\eqref{eq:eulerian_env}. Figure~\ref{fig:delta_crit_semi} shows the critical overdensity as a function of mass for different values of $\log_{10}|f_{R0}|$ for the Planck cosmology given in Table~\ref{tab:param_cosmo}~\cite[][hereafter Planck\,2018]{Planck2020}. For reference, Fig.~\ref{fig:delta_crit_semi} also shows the constant critical overdensity in a corresponding GR universe. Note that the critical overdensity in $f(R)$ gravity approaches the GR value for high halo masses, because $\frac{\Delta \xi}{\xi} \to 0$ for high masses and thus $F \to 0$. 
The critical overdensity in $f(R)$ gravity and in GR as well as the derivative of $\delta_{\mathrm{crit}}$ with respect to $\mathrm{dln}M$ are used to calculate the halo mass function in $f(R)$ gravity models.
\begin{table}
    \centering
    \caption{Fiducial cosmological parameters and their values for the flat Planck\,2018 cosmology \cite[][table\,2, column 4]{Planck2020}. $\Omega_{\mathrm{m}}$ is the matter density parameter, $\Omega_{\mathrm{b}}h^2$ is the reduced baryon density, $h$ is the Hubble parameter, $N_{\textnormal{eff}}$ is the effective number of relativistic species, $\sum m_\nu = \Omega_\nu h^2 \cdot 94\,\mathrm{eV}$ is the sum of neutrino masses, $A_s$ is the power spectrum amplitude parameter, $\sigma_8^{\rm GR}$ is the amplitude of the linear power spectrum at a scale $8\,h^{-1}\,\mathrm{Mpc}$ at redshift $z=0$ and $n_s$ is the scalar spectral index.}
    \begin{tabular}{lc}
    \hline
    \hline
    Parameter & Fiducial value \\
    \hline
    $\Omega_{\mathrm{m}}$ & 0.3166\\
    $\Omega_{\mathrm{b}}h^2$ & 0.02236 \\
    $h$ & 0.6727 \\
    $N_{\textnormal{eff}}$ & 3.046 \\
    $\Sigma m_\nu$ [eV] & 0.06 \\
    $\ln(10^{10}A_s)$ & $3.045$ \\
    $\sigma_8^{\rm GR}$ & 0.812\\
    $n_s$ & 0.9649\\
    \hline
    \hline
    \end{tabular}
    \label{tab:param_cosmo}
\end{table}

\subsection{\label{subsec:HMF} Halo mass function in \texorpdfstring{$f(R)$}{f(R)} gravity}
To study and forecast cosmological constraints with cluster surveys we have to understand the distribution of halos in mass and redshift, \ie the (differential) HMF. We model the HMF in an $f(R)$ gravity cosmology with two components. The first component is the GR HMF, and the second is the enhancement (or suppression) factor of the GR HMF due to $f(R)$ gravity. This factor accounts for the scale-dependent clustering of matter \cite{Shandera13,Cataneo14}. We adopt the GR halo mass function from~\cite{Tinker08}, which is given by
    \begin{equation}
        \label{eq:Tinker_HMF}
        \left. \frac{\dd n}{\dd \mathrm{ln} M} \right\vert_{\mathrm{T}} = - \frac{\bar \rho_{\mathrm{m}}}{2M} f(\sigma)_{\mathrm{T}} \frac{\dd \mathrm{ln} \sigma^2}{\dd \mathrm{ln} M}\, .
    \end{equation}
Here $f(\sigma)_{\mathrm{T}}$ is the multiplicity function \cite{Tinker08}
    \begin{equation}
        \label{eq:Tinker_multi_fct}
        f(\sigma)_{\mathrm{T}} = \tilde A \left [ \left(\frac{\sigma}{\tilde b}\right)^{-\tilde a} +1 \right]e^{-\frac{\tilde c}{\sigma^2}}\, ,
    \end{equation}
where $\tilde A,\ \tilde a,\ \tilde b$ and $\tilde c$ are parameters calibrated using $N$-body simulations~\cite[see][table 2]{Tinker08} and $\sigma=\sigma(M)$ is the variance of the overdensity on a mass scale $M$ in the corresponding GR cosmology. 

For the ratio of the HMF in $f(R)$ gravity to GR we use the Sheth~\&~Tormen halo mass function~\cite{Sheth1999}, which can account for the scale-dependent collapse through $\delta_{\mathrm{crit}}$. Furthermore, Refs.~\cite{Schmidt09,Lombriser13} showed that the Sheth~\&~Tormen prescription with modified collapse thresholds provides a good fit to the HMF in $N$-body simulations of $f(R)$ gravity. The Sheth~\&~Tormen HMF is given by 
    \begin{equation}
        \label{eq:ST_HMF}
        \left. \frac{\dd n}{\dd \mathrm{ln} M} \right\vert_{\mathrm{ST}} =  \frac{\bar \rho_{\mathrm{m}}}{M} f(\nu)_{\mathrm{ST}} \left[ \frac{\dd\mathrm{ln} \delta_{\mathrm{crit}}}{\dd\mathrm{ln} M} - \frac{1}{2} \frac{\dd\mathrm{ln} \sigma^2}{\dd\mathrm{ln} M} \right ]  \, ,
    \end{equation}
with $\nu = \delta_{\mathrm{crit}}/ \sigma$ the peak height and $f(\nu)_{\mathrm{ST}}$ the Sheth~\&~Tormen multiplicity function, which is parametrized as
    \begin{equation}
        \label{eq:ST_multi_fct}
        f(\nu)_{\mathrm{ST}} = A \sqrt{\frac{a\nu^2}{2\pi}} \left[1+(a\nu^2)^{-p}\right]e^{-\frac{a\nu^2}{2}}\, ,
    \end{equation}
where $A,\ a$, $p$ are free parameters. We adopt fitting formulae for these parameters of Ref.~\cite{Despali16}. This allows one to calculate the HMF for different halo overdensities $\Delta_{\mathrm{c}}$, \ie different halo definitions $M_{\Delta_{\mathrm{c}}} = 4/3 \pi \Delta_{\mathrm{c}} \bar \rho_{\mathrm{crit}} R^3 $. The fitting formulae are given by
    \begin{align}
        \label{eq:params_multi_fct}
        A &= -0.1362x + 0.3292 \, , \nonumber  \\
        a &= 0.4332x^2 + 0.2263x + 0.7665 \, ,\\
        p &= -0.1151x^2 + 0.2554x + 0.2488 \nonumber \, . 
    \end{align}
Here $x(z) = \log_{10} (\Delta/\Delta_{\mathrm{vir}}(z))$ and $\Delta_{\mathrm{vir}}(z)$ is the virial overdensity. 
In this model we use an effective virial overdensity from the corresponding GR cosmology which is given by \cite{Bryan1998}
    \begin{equation}
        \label{eq:vir_overdens}
        \Delta_{\mathrm{c, vir}}(z) = 18\pi^2 - 82\left[1-\Omega_{\mathrm{m}} (z)\right] - 39\left[1-\Omega_{\mathrm{m}} (z)\right]^2\, .
    \end{equation}

The difference between the Sheth~\&~Tormen HMF in $f(R)$ gravity and GR is completely encoded in the critical overdensity coming from Eq.~\eqref{eq:EoM_inner} and \eqref{eq:EoM_outer}. Therefore, we use an effective variance $\sigma=\sigma(M)$ and thus employ in both cases the GR value for the variance \cite{Lombriser13,Lombriser14}. Furthermore, if we include massive neutrinos, the shape of the HMF is more universal if only the cold dark matter and the baryon power spectrum is used to calculate the variance \cite{Ichiki12}. 
We adopt this also in the case of an $f(R)$ gravity cosmology to account for the effect of massive neutrinos, which assumes that neutrinos behave the same in modified gravity. This approach was followed in other studies ~\cite{Hagstotz19,Artis24}.
We vary also the mass of the neutrinos because there is a known degeneracy between the $f(R)$ gravity parameter and massive neutrinos \citep[see \eg][]{Motohashi13,Baldi14,Wright19}. In summary, our $f(R)$ HMF has the following form
    \begin{align}
        \label{eq:fR_HMF}
        \frac{\dd n}{\dd \mathrm{ln} M} &= \left. \frac{\dd n}{\dd \mathrm{ln} M} \right\vert_{\mathrm{T}} \times \mathcal{R}\, , \\
        \mathrm{with} \quad  \mathcal{R} &= \frac{\left. \frac{\dd n}{\dd \mathrm{ln} M} \right\vert_{\mathrm{ST},\ f(R)}}{\left. \frac{\dd n}{\dd \mathrm{ln} M} \right\vert_{\mathrm{ST},\ \mathrm{GR}}} \nonumber \, .
    \end{align}
    \begin{figure}
        \centering
        \includegraphics[width=\linewidth]{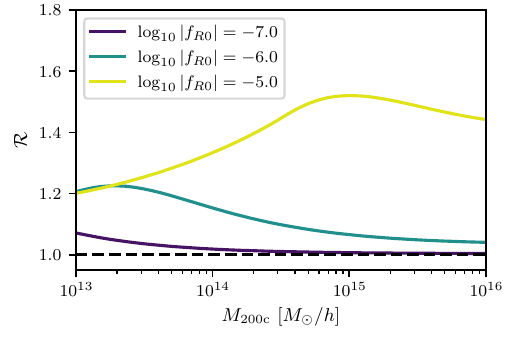}
        \vskip-0.25cm
        \caption{The ratio $\mathcal{R}$ of the $f(R)$ HMF and the GR Sheth~\&~Tormen HMF in Eq.~\eqref{eq:ST_HMF}, for different values of the $f(R)$ parameter $\log_{10}|f_{R0}|$ in colored lines.}
        \label{fig:HMF_fR}
    \end{figure} 

The difference between the HMF in an $f(R)$ gravity model to GR for different values of $\log_{10}|f_{R0}|$, \ie the ratio $\mathcal{R}$ from Eq.~\eqref{eq:fR_HMF}, is shown in Fig.~\ref{fig:HMF_fR}. 
Generally, $f(R)$ gravity enhances the growth of structure and thus $f(R)$ gravity predicts more clusters compared to a GR cosmology. 
As expected, models with larger $|f_{R0}|$ show larger differences in the HMF. Moreover, the shape of the enhancement depends on the strength of the $f(R)$ model. If the value of $|f_{R0}|$ is comparable to the cosmological potential $\Phi$ of massive dark matter halos (see Eq.~\eqref{eq:poisson_eq_2}), the fifth force is screened within these halos, so that the abundance of high mass halos is not increased substantially. This is the case for $|f_{R0}| \leq 10^{-6}$. In the case of $\log_{10}|f_{R0}| = -5$, a high enhancement for the most massive halos is seen, because for this the value of $f_{R0}$ even massive halos are at most partially screened.
Furthermore, Fig.~\ref{fig:HMF_fR} shows that the HMF enhancement is weaker for smaller $f_{R0}$. Consequently, distinguishing between such weak modified gravity models and GR using a cluster abundance analysis will be challenging (see Sec.~\ref{subsec:analysis_cl}).

\subsection{\label{subsec:emu} Emulating the critical overdensity quantities} 
The computational bottleneck in our HMF calculation is the computation of the critical overdensity in $f(R)$ gravity and GR as well as the derivative of $\delta_{\mathrm{crit}}$ with respect to $\mathrm{ln} M$. 
These quantities are obtained by solving a system of coupled differential equations for each mass and redshift individually in the range where we need the HMF and thus a lot of computations must be performed to obtain the HMF. 
To speed up the calculations of $\delta_{\mathrm{crit}}$, $\dd \delta_{\mathrm{crit}} / \dd \mathrm{ln} M$ and $\delta_{\mathrm{crit}, \mathrm{GR}}$ we create three separate emulators for these quantities

We choose to emulate the critical density and its derivative rather than the HMF directly, because a direct emulation of the HMF 
would require us to additionally sample baryon and neutrino density parameters, thereby making the emulation more complex. In the end, executing our emulators and computing the HMF is essentially as fast as emulating the HMF. 

We use Gaussian processes regression (GPR), a supervised learning method, to build the emulators. The emulators are trained on a data set that samples the desired parameter space and then verified on an independent validation dataset to assess the performance of the emulators. The parameter space of $\delta_{\mathrm{crit}}$ and $\dd \delta_{\mathrm{crit}} / \dd \mathrm{ln} M$ that we sample is seven-dimensional, with five cosmological parameters together with the halo mass $M$ and redshift $z$. This allows us to emulate the behavior of the coupled system of differential equations ~\eqref{eq:EoM_inner} and \eqref{eq:EoM_outer}. 
The ranges we choose for the parameters are
    \begin{align}
        \label{eq:range_params}
        z &\in [0, 2] \, , \nonumber \\
        \log_{10}M &\in [13, 16] \, , \nonumber \\
        \Omega_{\mathrm{m}} &\in [0.11, 0.4] \, , \nonumber \\ 
        h &\in [0.6, 0.82] \, ,  \\
        n_{\mathrm{s}} &\in [0.8, 1.1] \, , \nonumber \\
        \sigma_8^{\mathrm{GR}} &\in [0.6, 0.9] \, , \nonumber \\
        \log_{10}|f_{R0}| &\in [-7, -3] \nonumber \, .
    \end{align}
The ranges for the cosmological parameters are large enough to ensure that we are not hitting the boundaries in our likelihood sampling, see Sec.~\ref{subsec:analysis}. In the case of the halo mass and redshift, we choose ranges such that we cover the interval of masses and redshifts of clusters detectable with the surveys from SPT-3G and CMB-S4. 

The parameter space for $\delta_{\mathrm{crit}, \mathrm{GR}}$ is only two-dimensional with the parameters $\Omega_{\mathrm{m}}$ and $z$, because the spherical collapse equation, Eq.~\eqref{eq:EoM_outer}, for GR only depends on these two quantities. The ranges for these parameters are the same as for the two other emulators.

\begin{figure}
        \centering
        \includegraphics[width=\linewidth]{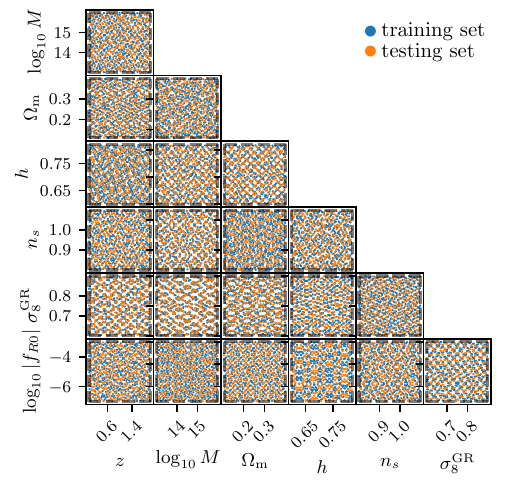}
        \vskip-0.25cm
        \caption{
        The 7-dimensional parameter space for the emulation of $\delta_{\mathrm{crit}}$ and $\dd \delta_{\mathrm{crit}} / \dd\mathrm{ln}  M$ drawn from a Sobol sequence algorithm. In total 1024 points are sampled. The training set (first half of the points) is marked in blue and the validation data are shown in orange. The ranges of the parameters are given in Eq.~\eqref{eq:range_params}.}
        \label{fig:sobol_space}
    \end{figure}

We sample the points in the parameter space with a Sobol sequence algorithm. This algorithm ensures that we sample the parameter space efficiently without duplicating values of any parameter in the sample, and thus it is a better choice than a uniform random or grid sampling algorithm. We sample $2^{10} = 1024$ points and use the first half for training and the second half for validating the emulator.
Both datasets are evenly distributed in the parameter space and are disjoint (\ie, no point in the parameter space can appear in both datasets), as guaranteed by the Sobol sequence algorithm. Figure~\ref{fig:sobol_space} shows the parameter space with training data in blue and the validation set in orange; the characteristic pattern for the Sobol sequence algorithm can be seen. 

The performance of the emulator is quantified by the normalized median absolute deviation (nMAD), which is a robust estimator of the scatter around the true value. 
The nMAD is a better quantity for the deviation from the average than the standard deviation if the dataset shows a large scatter. 

Our validation tests show emulator accuracy for $\delta_\mathrm{crit}$, $\dd\delta_\mathrm{crit}/\dd \mathrm{ln}M$ and $\delta_\mathrm{crit,GR}$ of 0.2\,\%, 22.9\,\% and 0.0002\,\%, respectively. 
Thus, the critical overdensity in $f(R)$ gravity and GR computed with the emulator is accurate at the sub-percent level. However, the logarithmic mass derivative of the critical overdensity can only be predicted at the $\sim25\,\%$ accuracy level by the emulator. 

We test also whether increasing the accuracy of the $\dd \delta_{\mathrm{crit}} / \dd \mathrm{ln} M$ emulator by reducing the dynamical range of the quantity and emulating $ \mathrm{ln}(\dd \delta_{\mathrm{crit}} / \dd \mathrm{ln} M)$ instead. Unfortunately, this approach does not increase the accuracy of the emulator. 
The relatively high inaccuracy can be explained by the inherent complexity of $\dd \delta_{\mathrm{crit}} / \dd \mathrm{ln} M$, which is the halo mass as a free parameter, along with a quantity that is emulated while involving differentiation with respect to the halo mass. 
We can reach a higher accuracy by either using more points in the training dataset or by further developing or replacing the GPR emulator used to emulate $\dd \delta_{\mathrm{crit}} / \dd \mathrm{ln} M$. 
However, our forecast depends on the accuracy of the HMF rather than the accuracy of the critical overdensity and derivative. 
As discussed in the next section, the error of the emulated HMF is mainly driven by the error of $\delta_{\mathrm{crit}}$, and the error of $\dd \delta_{\mathrm{crit}} / \dd \mathrm{ln} M$ does not have a big impact.

\subsection{\label{subsec:HMF_validation}HMF validation} 
Because we use the emulated critical overdensities in $f(R)$ gravity and GR as well as the logarithmic derivative with respect to mass, we examine the impact of inaccuracies in the emulation on the 
$f(R)$ HMF, Eq.~\eqref{eq:fR_HMF}. We compute the ratio $\mathcal{R}$, Eq.~\eqref{eq:fR_HMF}, for the points in the validation dataset once with the analytical result and compare them to the results using the emulated values.

    \begin{figure}
        \centering
        \includegraphics[width=\linewidth]{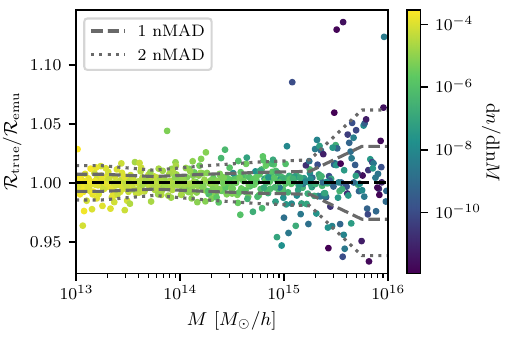}
        \vskip-0.25cm
        \caption{Ratio of the true HMF to the HMF calculated using the emulator values
        of $\delta_{\mathrm{crit}, \mathrm{GR}}$, $\delta_{\mathrm{crit}}$ and $\dd \delta_{\mathrm{crit}} / \dd \mathrm{ln} M$. The ratio is calculated for the validation dataset and shown as a function of halo mass; the validation dataset spans the full range of redshift, mass and cosmological parameters adopted for the emulation.
        The gray dashed and dotted lines represent the one and two $\sigma$ (actually nMAD) median absolute differences within six mass bins.}
        \label{fig:test_HMF}
    \end{figure} 
Figure~\ref{fig:test_HMF} shows the ratio $\mathcal{R}_{\mathrm{true}}/\mathcal{R}_{\mathrm{emu}}$ of the semi-analytical and the emulator results for the validation points as a function of mass. These validation points span the full range of redshift over which the emulators were trained. One can see that as the mass increases the scatter in the ratio $\mathcal{R}_{\mathrm{true}}/\mathcal{R}_{\mathrm{emu}}$ also increases, indicating a greater characteristic uncertainty in the HMF emulation for higher masses. Moreover, the accuracy of the HMF is worse for locations where the amplitude of the halo mass function is lower (as indicated by the color bar). 
We quantify the accuracy of the emulated HMF by the nMAD in six mass bins evenly spaced in logspace between $[10^{13}, 10^{16}]\,h^{-1}M_\odot$ and we obtain $\rm nMAD = (0.7, 0.5, 0.7, 0.8, 0.9, 3.0)\,\%$. Note that the nMAD is computed from the test set which is sampled in redshift, mass and cosmological parameters and thus averaged over the entire ranges of redshift and cosmological parameters.
As seen the nMAD in the last mass bin, \ie $[15.5, 16]\, \log[M/(h^{-1}M_\odot)]$, is significantly larger than that of the other bins. However, these ultra-high mass clusters are exceedingly rare in the Universe, and therefore the statistical uncertainties due to shot noise are larger than the two to five percent inaccuracies of the HMF in this regime. 

Furthermore, Fig.~\ref{fig:test_HMF} shows that the relatively large uncertainty in the derivative of the $f(R)$ critical overdensity with respect to $\mathrm{dln} M$ does not significantly affect the accuracy of the HMF. This is expected. First, the exponential form in the overdensity of the multiplicity function, Eq.~\eqref{eq:ST_multi_fct}, is sensitive to the absolute difference of the emulated and semi-analytical result of $\delta_{\mathrm{crit}}$. And second, $\delta_{\mathrm{crit}}$ has a much lower dynamical range than $\sigma$, \ie $\dd \delta_{\mathrm{crit}} / \dd\mathrm{ln}M$ is smaller than $\dd \sigma / \dd \mathrm{ln} M$ in the Sheth~\&~Tormen HMF, Eq.~\eqref{eq:ST_HMF} and therefore makes only a small contribution to the HMF.

The semi-analytical HMF has previously been shown to be consistent with simulations at the level of $20\,\%$ 
\cite{Lombriser13}. New results from the FORGE HMF emulator show agreement with the semi-analytical HMF at the $15\,\%$ level \cite{Ruan23}. Therefore, the inaccuracy in our HMF due to employing an emulator for the critical overdensities and logarithmic mass derivative is not a limiting factor in our analysis.

The calculation of the HMF with the emulator in a mass and redshift grid of twenty times twenty points and a Planck\,2018 cosmology with $\log_{10}|f_{R0}| = -6$ is more than three orders of magnitude faster in comparison to employing the semi-analytic model directly. In the case of a single core on a typical Linux computing cluster, this improvement reduces the computation time from half an hour to roughly 1.4\,seconds.

\section{\label{sec:surveys} Cosmological surveys}
In this analysis, we focus on the constraining power on $f(R)$ gravity of two different galaxy cluster samples when combined with weak gravitational lensing mass calibration. 
We describe here the two galaxy cluster samples, which are representative of SPT-3G and CMB-S4, and the ngWL weak lensing dataset needed for mass calibration, which could come from either the Euclid satellite or Rubin observatory. 

\subsection{\label{subsec:SZ_catalogs}tSZE-selected cluster catalogs}

This section describes the cluster survey specifics for SPT-3G and CMB-S4, which are used to forecast constraints on $f(R)$ gravity.
In modeling the cluster samples we 
follow the approach adopted in the recent SPT$\times$DES cluster cosmological analysis \cite{Bocquet24}. 
The massive galaxy clusters in the two samples are selected using their thermal SZE signature (tSZE). A matched filter tuned to detect the cluster tSZE is applied to the multi-frequency maps over a broad range of cluster core radii \cite{Melin06,Staniszewski09} to identify peaks. Given the noise in the maps and the amplitude of the tSZE peak, an associated detection significance $\hat\zeta$ is assigned. The approach produces a candidate list of tSZE selected clusters with an initial contamination fraction. The $\hat\zeta$ observable serves as a mass proxy for the cluster analysis as described in Section~\ref{sec:obs-mass-rel}. 

Each cluster candidate is then confirmed by applying a matched-filter technique to the optical and near-infrared (NIR) galaxy catalog at the location of each cluster \cite[e.g.,][]{Song12b,Bleem15,Klein23b}. 
In this process, each cluster is assigned a redshift $z$ and a cluster richness $\hat\lambda$, which corresponds to the color and position weighted number of passive galaxies within the cluster. For follow-up tools like the multi-component matched filter \cite[MCMF;][]{Klein18,Klein23b}, the chance that each candidate is a random superposition of a tSZE noise fluctuation and a physically unassociated optical/NIR system is also quantified. With this approach it is possible to exclude the likely contaminants from the tSZE selected list, producing a catalog with a specific targeted contamination (typically percent level).

Each confirmed cluster then has three observables used for the selection: tSZE significance $\hat\zeta$, richness $\hat\lambda$ and redshift $z$, and the sample for analysis is defined by the selection thresholds $\hat\zeta_{\rm min}$, $\hat\lambda_{\rm min}(z)$ and $z_\mathrm{min}<z<z_\mathrm{max}$. Here the redshift dependent richness selection comes from MCMF and is required to reduce the contamination in the tSZE candidate cluster list. The values of these thresholds depend on the cluster survey and are therefore specified in the next sections.

In our analysis, each cluster has a fourth observable, which is the weak lensing inferred mass, which we discuss further below.

\subsubsection{\label{subsec:SPT3G_data} SPT-3G cluster survey}
The South Pole Telescope (SPT) is a $10$\,m telescope operating in the microwave wavelength and is located near the Amundsen-Scott South Pole Station \cite{Carlstrom11}. 
Since 2018 the SPT has been equipped with a new, third-generation camera, SPT-3G \cite{Benson14}, which has been used to detect approximately an order of magnitude more clusters per sky solid angle than from the previous SPT surveys: SPT-SZ \cite{Bleem15,Klein23}, SPTpol~ECS \cite{Bleem20} and SPTpol~500d \cite{Bleem23}. 
In this work, we consider a $4,000\,\mathrm{deg}^2$ survey by SPT-3G that consists of two regions: a deep $1,500\,\mathrm{deg}^2$ field for which we model the selected cluster sample to be those with detection significance $\hat\zeta > 4.25$, and a shallower $2,500\,\mathrm{deg}^2$ field for which we adopt a selection $\hat\zeta > 5$.
Despite its smaller survey area, there are more detected clusters in the deep field because of its greater depth.

For the sample analyzed here, we assume that the full survey from SPT-3G is covered by the upcoming ngWL survey datasets similar to those that will be available from Euclid and Rubin. These datasets are crucial not only for the weak-lensing information to calibrate cluster masses, but also for the optical/NIR confirmation of SPT-3G cluster candidates in the range $z\in [0.25, 2]$.\footnote{We adopt the same lower redshift limit as adopted in previous cosmological analyses of SPT clusters \citep{Dehaan16, Bocquet19, Bocquet24}.} The selection threshold associated with the MCMF exclusion of contaminants is $\hat\lambda_\mathrm{min}(z)$, which we take to be the same as that used in the recent SPT$\times$DES analysis \cite{Bocquet23,Bocquet24}. These thresholds are chosen to lead to a contamination fraction in the final SPT-3G confirmed cluster sample at the percent level, where no explicit modeling of the contaminants will be required.

\subsubsection{\label{subsec:CMBS4mock} CMB-S4 cluster survey}
CMB-S4 is a future cosmic microwave background survey currently in the design and construction phase that will start operation at the end of this decade \cite{Abazajian19}. The survey will cover roughly 50\,\% of the sky, and the resulting mm-wave maps will be ideal for the detection of tens of thousands of clusters through their tSZE signatures, as shown in a recent forecast for CMB-S4 \cite{Raghunathan22}. 

Following this previous forecast, we adopt a cluster redshift range $z\in [0.1, 3]$ and a tSZE detection significance limit $\hat\zeta > 5$.
Note that this sample extends to lower redshift than the SPT samples because CMB-S4 has more frequency bands, enabling greater reliance on frequency filtering to remove the primary CMB anisotropies which affects cluster detection at low redshift if unmitigated.
We only consider clusters up to redshift $z = 2$ for the following two reasons. First, follow-up of clusters at $z > 2$ is not yet demonstrated and well understood. As previously noted, we expect the deep ngWL imaging datasets in the optical/NIR to be sufficient for follow-up to $z=2$. Second, we have chosen the Tinker HMF (Eq.~\eqref{eq:Tinker_HMF}) for the mock generation, and it is only calibrated at $z\lesssim 2$. 

We adopt a survey area corresponding to the overlap between CMB-S4 and the upcoming ngWL survey from the Euclid satellite. We use only the overlapping region to ensure that every confirmed CMB-S4 cluster will have available weak-lensing data. This is a conservative approach, because the HMF constraints from cluster samples do not require that each individual cluster have available weak-lensing data. With a Euclid-like footprint for our ngWL dataset the overlapping region is roughly $10,100\,\mathrm{deg}^2$
~\cite{Raghunathan22,Scaramella22}. The Rubin coverage would be even larger and would enhance the CMB-S4 sample relative to what we adopt here. For the optical/NIR selection $\hat\lambda_\mathrm{min}(z)$ from MCMF, we adopt also the selection thresholds recently used in the SPT$\times$DES analysis \cite{Bocquet23,Bocquet24}.

\subsection{\label{subsec:ngWL}Next-generation weak-lensing (ngWL) data} 
Next-generation weak-lensing surveys will collect an order of magnitude more lensing data than current WL surveys and will thus provide improved constraints on the parameters of the observable--mass relation relative to what is possible today.

In this analysis, we focus on ngWL data similar to that which we expect from the Euclid mission \cite{Laureijs11,Scaramella22}. Note that a similar analysis could be carried out for other ngWL surveys like the Legacy Survey of Space and Time conducted with the Vera C.\ Rubin Observatory \cite{Mandelbaum18}. Given the similarities of the weak-lensing datasets and photometric redshifts between the two surveys, we carry out a single analysis.

The most significant expected advantage in constraining power for Rubin compared to Euclid comes from the survey footprint. In the case of Rubin, the overlap with CMB-S4 is expected to be $\sim25\,\%$ larger than for Euclid, and thus constraints would be approximately $10\,\%$ tighter. For SPT-3G both Euclid and Rubin fully cover the planned survey region, and so there would be no expected significant differences in constraining power.

\begin{figure}
        \centering
        \includegraphics[width=\linewidth]{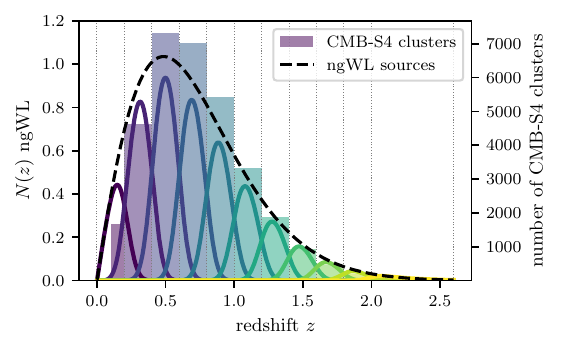}
        \vskip-0.25cm
        \caption{Normalized redshift distribution of the ngWL sources and the CMB-S4 cluster sample. The full ngWL source redshift distribution is split up into eleven tomographic bins. Colors are used to delineate redshifts.
        }
        \label{fig:source_red_distr}
    \end{figure}

We assume a lensing source density of $30\,\mathrm{arcmin}^{-2}$, a shape noise of 0.3, and a source redshift distribution with a median redshift $z_\mathrm{m}=0.9$ \cite{Laureijs11}. These characteristics are similar to the adopted goals for the ngWL surveys from Euclid and Rubin.
We assume an uncertainty in source redshifts of $\sigma_z=0.06$ \citep{Laureijs11}, and bin the lensing source galaxies into ten tomographic bins evenly spaced in redshift between $z=0$ and $z=2$, and add an eleventh bin in the redshift range 2.0--2.6.
Figure~\ref{fig:source_red_distr} shows the redshift distributions of the ngWL source galaxies and of the CMB-S4 clusters.
%

\section{\label{sec:obs-mass-rel} Observable-mass relations}

Galaxy cluster ensembles exhibit considerable regularity, shown first in tight observable--observable scaling relationships involving the X-ray sizes, temperatures and ICM masses \cite{Mohr1997,Mohr1999}
and then more recently in observable--mass scaling relations calibrated using weak lensing data \cite[e.g.][]{Hoekstra15,Mantz16,Chiu18,Chiu22}. Because galaxy clusters are typically identified and selected using their observable properties, while the HMF is expressed in terms of halo mass and redshift, scientific analyses of cluster samples usually rely on the existence of these observable--mass scaling relations. 
The analysis method we employ here is based on an empirical calibration of the observable--mass scaling relations using information from 
weak gravitational lensing based mass estimates as previously demonstrated with tSZE selected cluster samples from SPT \cite{Bocquet19,Bocquet23,Bocquet24}.

As mentioned in the introduction, while $f(R)$ gravity, in general, modifies the dynamics and hence also the tSZE signal, it does not affect gravitational lensing directly. To be precise, the lensing signal is rescaled by a factor of $(1+|f_{R0}|)^{-1}$ \cite{Sotiriou/Faraoni,Zhang07}, which for the range of $f_{R0}$ we consider is negligible, and we can thus assume that gravitational lensing is the same in $f(R)$ gravity and GR. Hence, we assume in the following that all modified gravity effects on the mass proxy are calibrated via the weak-lensing mass calibration.
In this work, we assume that $f(R)$ gravity does not strongly affect the halo shapes and that the GR-based weak-lensing mass calibration we will introduce in Section~\ref{sec:Mwl_mass} is valid.
A fully self-consistent approach will require $f(R)$ numerical simulations to properly account for the impact of halo shape changes on the inferred weak-lensing masses. We defer such an analysis to future work.

Here we summarize the observable--mass scaling relations relevant for our forecasts.

\subsection{\label{sec:zeta_mass}tSZE \texorpdfstring{$\zeta$}{zeta}--mass relation}
We assume the following relation for the mean {\it intrinsic} 
tSZE galaxy cluster detection significance $\zeta$
    \begin{equation}
    \begin{split}
        \label{eq:zeta_mass_rel}
        \langle \ln \zeta \rangle = \ln\asz &+ \bsz \ln\left (  \frac{M_{200\mathrm{c}}}{3 \times 10^{14}\, h^{-1} M_\odot} \right) \\
        &+ \csz\ln \left( \frac{E(z)}{E(0.6)} \right) \, .
    \end{split}
    \end{equation}
Here $\asz$, $\bsz$ and $\csz$ are the parameters corresponding to the normalization, mass trend and redshift trend of the relation and $E(z) = H(z)/H_0$.
We assume that the intrinsic detection significance $\zeta$ scatters around the mean relation in a log-normal fashion with a width described by $\sigmalnzeta$.

The intrinsic detection significance is given for a survey of a fiducial depth.
To account for field-to-field variation in the survey noise level and to use the cluster data from multiple fields of different depths to constrain the underlying $\zeta$--mass relation, we rescale the normalization $\asz$ of Eq.~\eqref{eq:zeta_mass_rel} using a factor $\gamma$; $\gamma=1$ corresponds to the average depth of a specific field within the SPT-SZ survey \citep{Dehaan16,Bleem20,Bleem23}.
Scaling up from the existing SPT-SZ and SPTpol cluster surveys, we adopt $\gamma = 3.5$ for the deep SPT-3G field and $\gamma = 1.5$ for the shallower field.
For CMB-S4 we adopt a scale factor $\gamma = 4$.
This factor is similar to the SPT-3G main field because at the depths of SPT-3G and CMB-S4 we expect the noise to be dominated by the cosmic infrared background. We find that with this scale factor, we recover similar numbers of clusters as in a previous forecast \cite{Raghunathan22b}.

The tSZE observed detection significance $\hat\zeta$ is related to the intrinsic detection significance $\zeta$ by a normal distribution\footnote{Note that in many past SPT analyses, the observed detection significance $\hat\zeta$ has been designated $\xi$. As we move to multi-observable analyses, we find it simplest to create observed-intrinsic pairs of observables using a single variable name with and without the hat.}
    \begin{equation}
        \label{eq:zeta_to_observed}
        P(\hat\zeta | \zeta) = \mathcal{N} \left( \sqrt{\zeta^2 +3}, 1 \right)  \, .
    \end{equation}
The normal distribution is due to the Gaussian noise in the survey maps, and the bias correction of 3 accounts for a noise bias introduced in the matched-filter search for peaks within three dimensions: location on the sky (2 parameters) and effective core radius of the tSZE signature \cite{Vanderlinde10}.
 
\subsection{\label{sec:lambda_mass}Cluster richness \texorpdfstring{$\lambda$}{lambda}--mass relation}

The mean intrinsic richness $\lambda$--mass scaling relation is described by a power law in mass and redshift
    \begin{equation}
        \label{eq:lambda_mass_rel}
        \begin{split}
        \langle \ln \lambda \rangle = \ln \alambda &+ \blambda \ln\left (  \frac{M_{200\mathrm{c}}}{3 \times 10^{14}\, h^{-1} M_\odot} \right) \\
        &+ \clambda\ln \left( \frac{1+z}{1.6} \right) \, ,
        \end{split}
    \end{equation}
with $\alambda$, $\blambda$ and $\clambda$ describing the normalization, mass trend and redshift trend. We assume that the cluster intrinsic richness scatters around this relation in a log-normal fashion described by its RMS variation $\sigmalnlambda$.

The observed richness $\hat\lambda$ is related to the intrinsic richness $ \lambda$ by
        \begin{equation}
        \label{eq:richness_to_observed}
        P(\hat \lambda | \lambda) = \mathcal{N} ( \lambda,  \sqrt{\lambda} )  \, ,
    \end{equation}
which posits the Poisson sampling noise associated with the realization of a particular number of observed galaxies $\hat\lambda$ in a galaxy cluster with an intrinsic richness $\lambda$. 
This expression is the Gaussian approximation to Poisson noise that we assume to be valid for $\lambda \gtrsim 10$.

\subsection{\label{sec:Mwl_mass}Weak lensing \texorpdfstring{$M_\mathrm{WL}$}{MWL}--mass relation}

The six scaling-relation parameters and the two intrinsic scatter parameters describing the $\zeta$--mass and $\lambda$--mass relations described above are calibrated using weak-lensing data, which have well characterized and controllable biases and uncertainties.
The WL observable for each cluster is the reduced tangential shear $\boldsymbol g_\mathrm{t}(R)$ or a collection of shear profiles, each associated with a different tomographic bin of weak lensing source galaxies. This observable is not employed in the cluster selection, but it is used to extract a weak-lensing halo mass $M_\mathrm{WL}$ by fitting the reduced shear profile to a Navarro-Frenk-White profile \cite[NFW;][]{Navarr01996} in the radial range $500\,h^{-1}\,\mathrm{kpc} < R < 3.2/(1+z)\,h^{-1}\,\mathrm{Mpc}$. The lower bound of the radial ranges ensures that the complex inner structure of the halo has minimal impact on our analysis. 

The derived WL mass $M_\mathrm{WL}$ will differ from the true halo mass, taken here to be $M_{200\mathrm{c}}$, due to model uncertainties and observational noise. 
The model uncertainties represent systematic uncertainties that do not average down with increasing number of weak lensing source galaxies. Therefore, we track these systematics by introducing an additional observable--mass scaling relation relating the WL and halo masses \cite{Becker11,Dietrich19,Grandis21}.
    \begin{equation}
        \label{eq:WL_mass_rel}
        \begin{split}
         &\left\langle \mathrm{ln} \left( \frac{M_{\mathrm{WL}}}{2 \times  10^{14}\, h^{-1} M_\odot}  \right)  \right\rangle =\bWL \\
        &\hspace{2.5cm}+ \bWLM \mathrm{ln} \left( \frac{M_{200\mathrm{c}}}{2 \times 10^{14}\, h^{-1} M_\odot} \right) \,,
        \end{split}
    \end{equation}
where \bWL\ is the logarithmic mass bias normalization and \bWLM\ is the mass trend in this bias. We discuss the redshift dependence in Section~\ref{subsec:mock_WL_data}.
In addition, the weak lensing mass $M_{\mathrm{WL}}$ exhibits a mass dependent log-normal scatter about the mean relation given by the variance $\sigma^2_{\ln \mathrm{WL}}$ 
    \begin{equation}
        \label{eq:WL_mass_var}
        \begin{split}
       &\mathrm{ln}\,\sigma_{\ln \mathrm{WL}}^2 =
       \sWL  \\
       &\hspace{2.cm} +\sWLM  \mathrm{ln}  \left( \frac{M_{200\mathrm{c}}}{2 \times 10^{14}\, h^{-1} M_\odot} \right)\, ,
        \end{split}
    \end{equation}
where \sWL\ is the normalization and \sWLM\ is the mass trend. The determination of the posterior distributions of the parameters of these two relations is discussed in Section~\ref{subsec:mock_WL_data}.

\section{\label{sec:mocks}Generating mock data}

The mock catalogs used in this work are created by drawing the data from our model, and thus we consider all statistical and systematic uncertainties. The model we have adopted is fully consistent with the recently analyzed SPT$\times$DES dataset \cite{Bocquet23,Bocquet24}, and therefore we expect it to be an excellent baseline description of these future cluster and weak-lensing datasets. 
In the following two sections, we describe in detail how we create the galaxy mock catalogs and the follow-up next-generation weak-lensing data.

\subsection{\label{subsec:mock_cluster_catalog} Mock cluster catalog}

Mock tSZE catalogs in a fiducial cosmology are created by first computing the HMF in the mass range $M_{200\mathrm{c}} \in [10^{13}, 10^{16}]\,h^{-1}M_\odot$ and in the redshift range of the given survey. 
We scale this HMF by the appropriate redshift-dependent volume, creating a function containing the expected number of halos of a given mass and redshift within mass and redshift bins:
    \begin{equation}
        \label{eq:number_cl_mock}
        \left<N(M,z)\right> \simeq  \frac{\dd n (\boldsymbol p,M,z)}{\dd M} 
        \frac{\dd V(\boldsymbol p, z)}{\dd z}
        \mathop{\dd M} \mathop{\dd z} \, ,
    \end{equation}
where the first factor is the HMF and the second factor is the survey solid angle $\Omega_{\rm s}$ dependent differential volume. 
We then create clusters of particular mass and redshift by drawing a Poisson realization of the expected number of halos within each mass and redshift bin. 

For each halo we assign a tSZE intrinsic detection significance $\zeta$ and an intrinsic optical richness $\lambda$ using the observable--mass relations Eqs.~\eqref{eq:zeta_mass_rel} and \eqref{eq:lambda_mass_rel} together with the associated log-normal scatter. 
For the mock catalog, the parameters of the observable--mass relations are fixed to the values $( \ln \asz, \bsz, \csz, \sigmalnzeta)= (0.96, 1.5, 0.5, 0.2)$ for the $\zeta$--mass relation and $( \ln \alambda, \blambda, \clambda, \sigmalnlambda) =(4.25, 1.0, 0.0, 0.2)$ for the $\lambda$--mass relation. The {\it observed} tSZE detection significance $\hat\zeta$ and the {\it observed} richness $\hat\lambda$ are then drawn using the intrinsic values and the measurement and sampling noise described in 
Eqs.~\eqref{eq:zeta_to_observed} and \eqref{eq:richness_to_observed}, respectively. 

Finally, we produce the cluster sample, modeling the selection by applying the appropriate lower thresholds in the tSZE detection significance $\hat\zeta_\mathrm{min}$ and observed richness $\hat\lambda_\mathrm{min}(z)$ within the redshift range adopted for each survey. The threshold value in tSZE detection significance and the redshift range for each survey are presented in Secs.~\ref{subsec:SPT3G_data} and \ref{subsec:CMBS4mock}. 

The minimum richness threshold as a function of redshift adopted here is modeled on the MCMF follow-up method \cite{Klein23b} and follows from the recent SPT$\times$DES analysis~\cite{Bocquet23,Bocquet24}.
Because that analysis only contains the minimum richness threshold up to $z = 1.79$, we extrapolate using $\hat\lambda_\mathrm{min}(z)$ equal to the value of $\hat\lambda_{\mathrm{min}}$ at $z=1.79$ for all higher redshifts considered in these forecasts.

We check the sensitivity of the cluster sample to different richness cuts by assuming a constant $\hat\lambda_\mathrm{min}$ for all redshifts with values of 10, 5 and 1. The number of clusters only varies by a few percent among these samples, similar to the variation we see when creating mocks with different random seeds. This is an indication that the tSZE significance selection threshold $\hat\zeta_\mathrm{min}$ is dominating the selection, and that the richness threshold has only a weak impact on the mock sampling and therefore the exact values we adopt are not important for our forecasts.

\begin{figure}
        \centering
        \includegraphics[width=\linewidth]{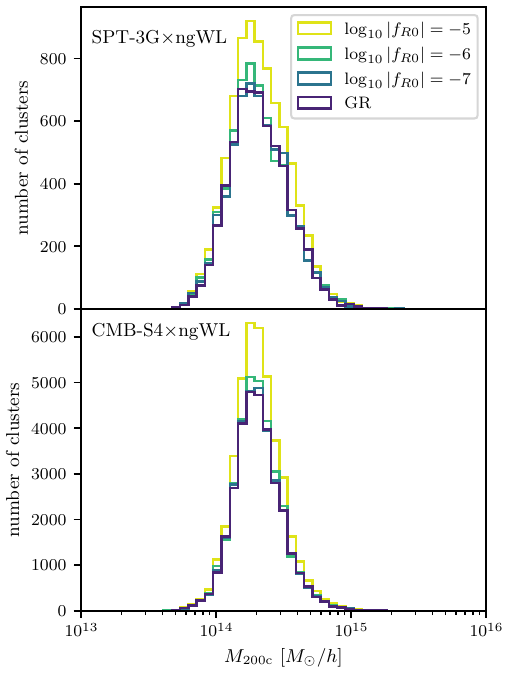}
        \vskip-0.25cm
        \caption{Mass distribution of the SPT-3G$\times$ngWL (above) and CMB-S4$\times$ngWL (below) mock catalogs for different $f(R)$ gravity models in yellow, light green and dark green as well as for the GR Planck\,2018 cosmology in dark violet.}
        \label{fig:mock_cat_dif_fR}
    \end{figure}

To create $f(R)$ gravity mocks we use the HMF described in Eq.~\eqref{eq:fR_HMF}. Here the number of clusters depends also on the strength of the gravity modification, with more clusters being obtained for stronger $f(R)$ models, as demonstrated in the enhancement of the HMF visible in Fig.~\ref{fig:HMF_fR}. Figure~\ref{fig:mock_cat_dif_fR} shows the distribution and abundance of clusters in mass for different values of $\log_{10}|f_{R0}|$ for our two different mock surveys, which we refer to as SPT-3G$\times$ngWL and CMB-S4$\times$ngWL.

As we can see from Fig.~\ref{fig:mock_cat_dif_fR}, cluster catalogs with a GR cosmology and $f(R)$ gravity with $\log_{10}|f_{R0}| = -7$ have close to the same number of clusters. This is in agreement with the difference in the HMF of these two models, which is only a few percent in the lower mass range (see Fig.~\ref{fig:HMF_fR}). Therefore, we assume in our analysis that an $f(R)$ gravity model with $\log_{10}|f_{R0}| = -7$ is indistinguishable from GR, and we choose the lower bound of $\log_{10}|f_{R0}|$ in the emulators described in Sec.~\ref{subsec:emu} to be $\log_{10}|f_{R0}| = -7$.

\subsection{\label{subsec:mock_WL_data} Mock ngWL data}
\begingroup
\renewcommand{\arraystretch}{1.25}
\begin{table}
    \centering
    \caption{Priors on the weak-lensing mass to halo mass relation parameters and parameter uncertainties (see Eqs.~\eqref{eq:WL_mass_rel} and \eqref{eq:WL_mass_var}). The mean values are adopted from a recent SPT$\times$DES analysis~\cite{Bocquet23}, and the uncertainties are assumed to be two times smaller. 
    }
    \begin{tabular}{cccc}
    \hline
    \hline
    Parameter & Prior & Uncertainty & Prior \\
    \hline
    $\ln M_{\mathrm{WL}, 0}$ & -0.050 & $\Delta\bWL$ & 0.010  \\
    $\bWLM$ & 1.029  & $\Delta \bWLM$ & 0.009  \\
    $\sWL$ & -3.100 & $\Delta \sWL$ & 0.120  \\
    $\sWLM$ & -0.226  & $\Delta \sWLM$ & 0.300 \\
    \hline
    \hline
    \end{tabular}
    \label{tab:WL_params}
\end{table}
\endgroup
We create mock cluster lensing data following the approach taken in the recent SPT$\times$DES analysis \citep{Bocquet23}. The first step is to use Eqs.~\eqref{eq:WL_mass_rel} and \eqref{eq:WL_mass_var} to assign a weak-lensing mass $M_{\mathrm{WL}}$ to each tSZE selected cluster in the mock sample. Doing so ensures that we include the fact that the weak-lensing mass is not the same as the halo mass. We then use this weak-lensing mass and the corresponding NFW profile 
to create a tangential shear profile $\boldsymbol g_\mathrm{t}(R)$ for each background tomographic bin (a detailed computation is shown in Refs.~\cite{Bocquet23,Bocquet24} on which our analysis is based). 
Source galaxy number densities and their redshift distribution are presented in Section~\ref{subsec:ngWL}.

In addition, we add cluster member contamination, consistent with the recent measurement in DES data \cite{Bocquet23}. This process produces realistically noisy and biased tangential shear profiles for each cluster. Specifically, these shear profiles include all the known systematic and stochastic effects needed to model cluster shear profiles in DES data. 

The parameters describing the $M_\mathrm{WL}$-$M_{200\mathrm{c}}$ relations (Eqs.~\eqref{eq:WL_mass_rel} and \eqref{eq:WL_mass_var}) 
are derived through the application of our mass measurement technique to hydrodynamical simulations of clusters. Following a previous work \cite{Grandis21}, the weak-lensing mass is extracted from hydrodynamical simulations of clusters, and associated $N$-body simulations are used to determine the corresponding true halo masses. 
Given both masses, the model for the $M_{\rm WL}$--$M_{200\mathrm{c}}$ relation and its scatter is calibrated.
The true halo masses adopted need to match those that are used to calibrate the HMF and its dependence on cosmology. In general, systematic uncertainties in the shear and photometric redshift estimates of the source galaxies can also play an important role in this weak-lensing to halo mass relation \cite{Bocquet23}.

Crucial in this process of parameter estimation for the weak-lensing to halo mass relation is that there are remaining uncertainties represented by the parameter posteriors. These represent an effective systematic error floor in our ability to estimate cluster halo masses using weak-lensing data.
The uncertainties on the bias and scatter parameters, $\Delta\bWL$, $\Delta \bWLM$, $\Delta \sWL$ and $\Delta \sWLM$, are then marginalized over during the mass calibration analysis to properly include the impact of weak lensing systematic uncertainties on the cluster observable--mass relations.

Table~\ref{tab:WL_params} contains the parameter values of the weak-lensing mass to halo mass relation that are adopted for the creation of the mock shear profiles and for the analysis results presented below. We use as a baseline the results from the DES weak-lensing calibration adopted in the recent SPT$\times$DES analysis~\cite{Bocquet23}.
We assume no redshift dependence in any of the $M_{\mathrm{WL}}$--$M_{200\mathrm{c}}$ scaling relation parameters in our analysis here because that dependence is largely driven by uncertainties in the photometric redshift biases in the DES shear catalog. The photometric redshift requirements for both Euclid and Rubin are so tight that related systematic uncertainties will be subdominant in an ngWL mass calibration analyses.

Compared to the current state-of-the-art, the ngWL datasets will include improvements in photometric redshift biases, shear measurement biases, and the understanding of the baryonic physics impact on the cluster mass profiles.
Therefore, we assume that the uncertainties on the $M_{\mathrm{WL}}$--$M_{200\mathrm{c}}$ parameters will improve by a factor of two compared to DES. We view this as a relatively conservative choice for the improvement of the WL uncertainties because in addition to better characterization of the cluster population through simulations, we can alter our analysis approach to reduce sensitivity to certain systematics (e.g., avoiding more of the cluster core where the baryonic effects are most important and selecting background source galaxies conservatively to dramatically reduce the cluster member contamination).

In Appendix~\ref{app:better_WL_uncertainties}, we discuss the impact of adopting more optimistic parameter uncertainties for ngWL surveys that are a factor of 10 smaller than those in the recent DES analysis; interestingly, moving from 2 times to 10 times reduction in systematic uncertainties has a minimal impact on the ngWL mass calibration. 
This suggests that at the level of weak lensing systematics adopted in our ngWL forecast, our observable--mass relation parameter posteriors are shape noise dominated.

\section{\label{sec:likel_analysis}Likelihood and Analysis}
The analysis method we employ in this work is based on the state-of-the-art SPT$\times$DES cosmological analysis \cite{Bocquet23,Bocquet24}. We summarize in this section the abundance and mass calibration likelihood and how we analyze our mock data for the SPT-3G$\times$ngWL and CMB-S4$\times$ngWL sample.

\begin{widetext}

\subsection{\label{subsec:likelohood} Abundance and mass calibration likelihood}
The multi-observable 
likelihood used in this forecast is given by
    \begin{equation}
      \label{eq:poisson_likelihood}
        \ln \mathcal L(\boldsymbol p) = \sum_i \ln \int_{\hat\lambda_\mathrm{min}}^\infty \dd\hat\lambda \frac{\dd^3 N(\boldsymbol p)}{\dd\hat\zeta \, \dd\hat\lambda \, \dd z} \Big|_{\hat\zeta_i, z_i}
        - \int_{z_\mathrm{min}}^{z_\mathrm{max}} \dd z \int_{\hat\zeta_\mathrm{min}}^\infty \dd\hat\zeta \int_{\hat\lambda_\mathrm{min}}^\infty \dd\hat\lambda \frac{\dd^3 N(\boldsymbol p)}{\dd\hat\zeta \, \dd\hat\lambda \, \dd z } 
        + \sum_i \ln\frac{\frac{\dd^4 N(\boldsymbol p)}{\dd\hat\zeta \, \dd\hat\lambda \, \dd \boldsymbol g_\mathrm{t} \dd z}
        \Big|_{\hat\zeta_i, \hat\lambda_i, \boldsymbol{g}_{\mathrm{t},i}, z_i}}
        {\int_{\hat\lambda_\mathrm{min}}^\infty \dd\hat\lambda \frac{\dd^3 N(\boldsymbol p)}{\dd\hat\zeta \, \dd\hat\lambda \, \dd z}
        \Big|_{\hat\zeta_i, z_i}}
        + \mathrm{const.} 
    \end{equation}
where both sums run over all clusters $i$.\footnote{More details and motivation of the likelihood approach can be found in Ref.~\cite{Bocquet23}.}
The vector $\boldsymbol p$ contains the cosmological and scaling relation parameters, and the observables are tSZE detection significance $\hat\zeta$, richness $\hat\lambda$, tangential shear profile $\boldsymbol{g}_{\mathrm{t}}$ and redshift $z$. The first two terms of the above equation represent the Poisson likelihood associated with the cluster abundance, which is independent of the weak-lensing data, while the last term represents the information from the mass calibration with the WL data.

The differential cluster number $\frac{\dd^3N}{ \dd\,\mathrm{obs}}$ that appears in the first two terms is the differential halo observable function HOF in the observable space $\hat\zeta - \hat\lambda - z$ 

    \begin{equation}
        \label{eq:HOF_with_3_obs}
        \begin{split}
        \frac{\dd^3 N (\boldsymbol p)}{\dd\hat\zeta \, \dd\hat\lambda \, \dd z } =
        \int \dd\Omega_\mathrm{s} 
        \iiint  \dd M\, \dd\lambda\, \dd\zeta\, P(\hat\zeta|\zeta) P(\hat\lambda|\lambda)P(\zeta, \lambda |M,z,\boldsymbol p) 
          \frac{\dd^2 N (M, z, \boldsymbol p)}{\dd M \, \dd V} 
         \frac{\dd^2 V(z,\boldsymbol p)}{\dd z \, \dd\Omega_\mathrm{s}} 
         \, .
        \end{split}
    \end{equation}
Here $P(\hat\zeta|\zeta)$ and $P(\hat\lambda|\lambda)$ follow from Eqs.~\eqref{eq:zeta_to_observed} and \eqref{eq:richness_to_observed}, respectively, whereas $P(\zeta, \lambda |M,z,\boldsymbol p)$ is obtained from Eqs.~\eqref{eq:zeta_mass_rel} and \eqref{eq:lambda_mass_rel}, $\Omega_\mathrm{s}$ is the survey solid angle, and the factors 
$\frac{\dd^2 N (M, z, \boldsymbol p)}{\dd M \dd z}$ and $\frac{\dd^2 V (z, \boldsymbol p)}{\dd z \dd\Omega_\mathrm{s}}$ 
are the HMF and the differential volume element for the corresponding cosmology.
The other differential cluster number in the observable space $\hat\zeta - \hat\lambda - \boldsymbol g_\mathrm{t} - z$ is given by
%
    \begin{equation} 
        \label{eq:HOF_with_4_obs}
        \begin{split}
        \frac{\dd^4 N(\boldsymbol p)}{\dd\hat\zeta \, \dd\hat\lambda \, \dd \boldsymbol g_\mathrm{t} \dd z } = 
        \int \dd\Omega_\mathrm{s}
        \iiiint  \dd M\, \dd\zeta\, \dd\lambda\, \dd M_\mathrm{WL}\, &
        P(\boldsymbol g_\mathrm{t}|M_\mathrm{WL}, \boldsymbol p)
        P(\hat\zeta|\zeta) P(\hat\lambda|\lambda) 
        P(\zeta, \lambda, M_\mathrm{WL} |M,z,\boldsymbol p) 
          \frac{\dd^2 N (M, z, \boldsymbol p)}{\dd M \, \dd V} 
         \frac{\dd^2 V(z,\boldsymbol p)}{\dd z \, \dd\Omega_\mathrm{s}} 
        \, ,
        \end{split}        
    \end{equation}
%
Where $P(\zeta, \lambda, M_\mathrm{WL} |M,z,\boldsymbol p)$ follows from Eqs.~\eqref{eq:zeta_mass_rel}, \eqref{eq:lambda_mass_rel} and \eqref{eq:WL_mass_rel} and $P(\boldsymbol g_\mathrm{t}|M_\mathrm{WL}, \boldsymbol p)$ is given by the product of Gaussian probabilities in each radial bin $i$ of the tangential shear profiles (see Sec.~\ref{subsec:mock_WL_data})
    \begin{equation}
    \begin{split}
        \label{eq:lensing_likelihood}
        P(\boldsymbol g_\mathrm{t}|M_\mathrm{WL}, \boldsymbol p) = \prod_i \left(\sqrt{2\pi}\Delta g_{\mathrm{t},i} \right)^{-1} 
        \exp \left[ -\frac12 \left(\frac{g_{\mathrm{t},i} - g_{\mathrm{t},i}(M_\mathrm{WL}, \boldsymbol p)}{\Delta g_{\mathrm{t},i}}\right)^2 \right] \, ,
    \end{split}
    \end{equation}
%
with the shape noise $\Delta g_{\mathrm{t},i}$.

\end{widetext}

\subsection{\label{subsec:analysis} Analysis}

The analysis presented in this work is done with {\textsc{CosmoSIS}}\footnote{\url{https://cosmosis.readthedocs.io/}}\cite{Zuntz14} using the Multinest and Nautilus samplers \cite{Feroz09, Lange23}. 
In our analysis we separate the cluster abundance and mass calibration elements into independent MCMC chains. 
The mass calibration likelihood is first used to quantify posterior distributions on the parameters of the observable--mass relations at a fixed cosmology. Thereafter, we adopt these parameter constraints as Gaussian priors on the parameters of the observable--mass relation parameters when evaluating cosmological constraints with the cluster abundance likelihood. The advantage of this approach is that it simplifies and dramatically speeds up the likelihood calculation. 
The downside is that the cosmological sensitivity of the mass calibration element (primarily due to $\Omega_\mathrm{m}$ sensitivity of the distance-redshift relation) is not correctly captured. As discussed in 
Appendix~\ref{app:cosmos_inteo_mass_calibration}, this separation of the analysis has little impact on our forecast cosmological or scaling relation parameter posteriors.

\subsubsection{\label{subsec:analysis_mass_cal}Mass calibration}
The main bottleneck in our mass calibration likelihood is the four-dimensional convolution integral from Eq.~\eqref{eq:HOF_with_4_obs}, which has to be evaluated for each cluster in our current implementation. In the calculation of this integral, we use an efficient Monte-Carlo integration method \cite{Bocquet23}; however, the total number of clusters is so large that it is not possible to complete a mass calibration chain for either mock cluster sample in a reasonable amount of time.
One approach to speed up the calculation would be to adopt a stacked analysis of cluster shear profiles \citep{Singh24}, but for this analysis, we use an approximate approach employing the likelihood in Eq.~\eqref{eq:poisson_likelihood} to infer the observable--mass relation parameter posteriors.
In this approach, we use 1000 randomly selected clusters from each tSZE survey to calculate the mass calibration likelihood and then scale up the weights of each selected cluster by a factor $\alpha$ that accounts for the number of missing clusters from the full sample. To up-weight the clusters we reduce the shape noise $\Delta g_{\mathrm{t}}$ by a factor of $1/\sqrt{\alpha}$ where $\alpha$ is the rescaling factor to the total number of clusters $N$, \ie $N = \alpha 1000$. 
This impacts the probability $P(\boldsymbol g_\mathrm{t}|M_\mathrm{WL}, \boldsymbol p)$ (defined in Eq.~\eqref{eq:lensing_likelihood}) of observing a particular reduced shear profile given the WL mass, appropriately rescaling the third term in the likelihood in Eq.~\eqref{eq:poisson_likelihood} to account for the full sample of clusters in the survey.

Because we assume that the mass calibration analysis is independent of the underlying cosmology,
we adopt fixed cosmological parameters that equal those used in generating the mocks and uniform priors on the eight observable--mass scaling relation parameters. The ranges for our flat priors are chosen such that they are larger than the $5\sigma$ results of the observable mass relations from previous analyses \cite{Bocquet19,Saro15}.
Importantly, we account for the uncertainties in the weak-lensing mass to halo mass relation (see Eqs.~\eqref{eq:WL_mass_rel} and \eqref{eq:WL_mass_var})
by adopting mean parameter values and parameter uncertainties as listed in Table~\ref{tab:WL_params} and then marginalizing over those uncertainties.
\begingroup
\renewcommand{\arraystretch}{1.25}
\begin{table}
    \centering    
    \caption{Priors on the parameters of our cluster abundance analysis for the SPT-3G$\times$ngWL (second column) and CMB-S4$\times$ngWL (third column) cluster samples. For the scaling relation parameters, the mean values are the same as the mock inputs, and the uncertainties are sampled from the posteriors of the corresponding mass calibration MCMC chain (see Sec.~\ref{subsec:analysis_mass_cal}). The prior on $\Omega_\nu h^2$ corresponds to a prior on the sum of neutrino masses $\sum m_\nu \sim \mathcal{U}(0, 0.6)~\mathrm{eV}$ }
    \begin{tabular}{lcc}
    \hline
    \hline
    Parameter \quad & SPT-3G$\times$ngWL \quad & CMB-S4$\times$ngWL\\
    \hline
    \multicolumn{3}{l}{Cosmology} \\
    $\Omega_{\mathrm{m}}$ & $\mathcal{U}(0.232, 0.4)$ & $\mathcal{U}(0.232, 0.4)$ \\
    $\Omega_{\nu}h^2$ & $\mathcal{U}(0, 0.00644)$  &  $\mathcal{U}(0, 0.00644)$   \\    
    $h$ & \quad \quad $\mathcal{U}(0.6, 0.8)$ \quad \quad & $\mathcal{U}(0.6, 0.8)$  \\
    $\ln(10^{10}A_s)$ & $\mathcal{U}(1, 4)$ & $\mathcal{U}(1, 4)$  \\
     $n_s$ & $\mathcal{U}(0.94, 1.)$ & $\mathcal{U}(0.94, 1.)$ \\
    $\log_{10}|f_{R0}|$ & $\mathcal{U}(-7, -3)$ & $\mathcal{U}(-7, -3)$\\
    \hline
    \multicolumn{3}{l}{tSZE $\zeta$-mass relation (Eqs.~\eqref{eq:zeta_mass_rel} and \eqref{eq:zeta_to_observed})} \\
    $\ln\asz$ & $\mathcal{N}(0.96, 0.03)$ & $\mathcal{N}(0.960, 0.021)$ \\
    $\bsz$ & $\mathcal{N}(1.50, 0.04)$  & $\mathcal{N}(1.50, 0.03)$  \\
    $\csz$ & $\mathcal{N}(0.50, 0.17)$ & $\mathcal{N}(0.50, 0.09)$  \\
    $\sigmalnzeta$ & $\mathcal{N}(0.200, 0.026)$  & $\mathcal{N}(0.200, 0.018)$ \\
    \hline
    \multicolumn{3}{l}{Richness $\lambda$-mass relation (Eqs.~\eqref{eq:lambda_mass_rel} and \eqref{eq:richness_to_observed})} \\ 
    $\ln \alambda$ & $\mathcal{N}(4.250, 0.019)$ & $\mathcal{N}(4.250, 0.014)$ \\
    $\blambda$ & $\mathcal{N}(1.00, 0.03)$  & $\mathcal{N}(1.000, 0.025)$   \\
    $\clambda$ & $\mathcal{N}(0.00, 0.16)$   & $\mathcal{N}(0.00, 0.06)$   \\
    $\sigmalnlambda$ & $\mathcal{N}(0.200, 0.012)$ & $\mathcal{N}(0.200, 0.008)$  \\
    \hline
    \hline
    \end{tabular} \label{tab:priors_cluster_abundance}
\end{table}
\endgroup

\subsubsection{\label{subsec:analysis_cl}Cluster abundance}

We adopt the posterior parameter distributions for the observable--mass scaling relation parameters that are listed in Table~\ref{tab:priors_cluster_abundance} as priors for the abundance analysis. 
In the cluster abundance likelihood analysis (see Eq.~\eqref{eq:poisson_likelihood}) we vary the cosmological parameters $\Omega_{\mathrm{m}},\ \Omega_{\nu}h^2,\ h,\ \mathrm{ln}(10^{10}A_s),\ n_s$, and $\log_{10}|f_{R0}|$. Note that $\sigma_8^{\rm GR}$ is a derived parameter. It is important to note here that the $\sigma_8^{\rm GR}$ quantity is the GR value as we use the linear power spectrum in the corresponding GR cosmology to calculate the HMF (see Sec.~\ref{subsec:HMF}). 

The goal of this work is to test the constraining power from cluster data with weak-lensing mass calibration alone, and thus we adopt flat priors on all cosmological parameters. 
However, as mentioned in Sec.~\ref{subsec:HMF}, it is difficult to distinguish between very weak modified gravity models and GR. 
Thus, efficiently sampling the likelihood for these weak models is challenging. 
Therefore, we combine the cluster abundance dataset with the primary CMB Planck\,2018 (TT, TE, EE) data for the analysis of GR and the $\log_{10}|f_{R0}| = -7$ mock catalogs.
This allows us to achieve convergence on a significantly more reasonable timescale. Note that we do not include future primary CMB results in this work. 
We account for the modified gravity sensitivity of the Planck\,2018 data by using the CMB power spectrum from $f(R)$ gravity in this part of the likelihood.
The cluster abundance data and Planck\,2018 data can be combined, because the mock catalogs are generated for a Planck\,2018 cosmology. Thus, there is no tension between the datasets. The Planck\,2018 likelihood is publicly available and implemented in \textsc{Cosmosis}. For the joint analysis we multiply the CMB likelihood with the cluster abundance likelihood.

Note that constraints on $f(R)$ gravity from Planck\,2018 primary CMB are of the order of $\log_{10}|f_{R0}| \lesssim -3$ \cite{Planck2020,Planck15}, and thus our constraints are dominated by the cluster abundance. The primary CMB data helps to anchor the standard cosmological parameters.

Table~\ref{tab:priors_cluster_abundance} summarizes the priors on the cosmological and scaling relation parameters for the SPT-3G$\times$ngWL and the CMB-S4$\times$ngWL forecasts. Note that we do not apply the lower bound from oscillation experiments for the massive neutrinos. 
The neutrino mass can only be constrained when combining with primary CMB data, and thus the upper bound for the neutrino prior is chosen based on the Planck\,2018 results.

Note that when the mock input value of the $f(R)$ parameter lies on the lower boundary of the $\log_{10}|f_{R0}|$ prior, the credibility limits are computed from the lower boundary, and we present only upper bounds. We apply this to the mock catalog with $\log_{10}|f_{R0}| = -7$ and the GR mock. Furthermore, the GR limit with $|f_{R0}| = 0$ is not reachable in a log prior and thus would add an infinitely large volume below our lower bound. To avoid dependence of the parameter upper limits on the choice of the prior lower boundary, we calculate the upper bounds in linear $|f_{R0}|$ space by transforming the parameter space from logarithmic to linear. In linear space the volume between $0$ and $10^{-7}$ is negligible.

\section{\label{sec:results} Results}
In this section, we summarize and discuss our results from the mass calibration and cluster abundance analyses for the future SPT-3G$\times$ngWL and CMB-S4$\times$ngWL datasets.

\subsection{\label{subsec:mass_calibration_constraints} ngWL mass calibration}
We expect that SPT will detect around 6,000 clusters in a Planck\,2018 cosmology with the SPT-3G camera. Applying the up-weighting approach described in Sec.~\ref{subsec:analysis_mass_cal} with $\alpha = 6$ for the mass calibration, we obtain constraints for the uncertainties of the observable--mass scaling relation parameters as shown in the second column of Table~\ref{tab:priors_cluster_abundance}. For the $\sim32,000$ CMB-S4 clusters, we use the up-weighting approach with $\alpha = 32$. The resulting parameter constraints for CMB-S4$\times$ngWL are shown in the third column of Table~\ref{tab:priors_cluster_abundance}.
We note that the mean recovered parameters are in agreement with the mock inputs within the uncertainties, confirming the validity of our analysis pipeline.

The tighter posteriors for the CMB-S4$\times$ngWL analysis are expected because SPT-3G$\times$ngWL has roughly five times fewer clusters (and therefore much less ngWL information).
Furthermore, the CMB-S4 dataset includes clusters at $z>0.1$, while for the SPT-3G dataset we adopt $z>0.25$ (see Sec.~\ref{subsec:SZ_catalogs}). We see the largest improvement in the uncertainty on the $\csz$ and $\clambda$ parameters because these two parameters model the redshift dependence of the observables at fixed mass.

We use the uncertainties from the mass calibration analyses presented here in the cluster abundance analyses presented in the next section.
Note that in the cluster abundance analyses we set the mean parameter values to be equal to the input values of the mocks. In addition, note that we have ignored the impact of modified gravity on the halo profile in our mass calibration analysis for this forecast, but it could be included in the analysis of the real datasets.

\subsection{\label{sec:cosmo_analysis}Cosmology constraints}
We now present the main results of the paper: the constraints on $f(R)$ modified gravity models for the future datasets SPT-3G$\times$ngWL and CMB-S4$\times$ngWL. We analyze four different models
with our cluster abundance pipeline:
values of $\log_{10}|f_{R0}| = \{ -7, -6, -5 \}$, denoted as F7, F6, and F5, respectively, as well as GR. The F7 model is assumed to be equivalent to GR in our framework (see Sec.~\ref{sec:mocks}), and the comparison of the F7 and GR datasets allows us to test this assumption.

In the sections that follow we compare our results with the $95\,\%$ upper bound constraint of $\log_{10}|f_{R0}| < -4.79$ found from clusters of galaxies combined with CMB data, CMB lensing, baryon acoustic oscillations (BAO) and type Ia supernova data \cite{Cataneo14}.
This constraint is also comparable to the upper end of the range of $f_{R0}$ constraints obtained from the Planck tSZE cluster sample~\cite{Peirone17}, where it is emphasized that uncertainties in the HMF prediction they adopted contribute significantly to their final constraint on $f_{R0}$. As discussed in previous sections, we have attempted to fold HMF uncertainties into our constraints.
\begin{figure*}
        \centering
        \includegraphics[width=\textwidth]{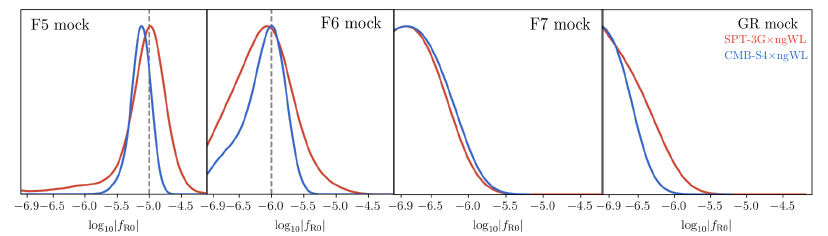}
        \vskip-0.25cm
        \caption{$\log_{10}|f_{R0}|$ posterior distributions from the cluster abundance analysis for SPT-3G$\times$ngWL (red) and CMB-S4$\times$ngWL (blue) datasets. For each survey one of the three independent mocks was chosen randomly. For the analyses of the GR and F7 models, the Planck\,2018 anisotropy data were included.}
        \label{fig:posterior_fR0}
\end{figure*}

\subsubsection{\label{subsec:cosmo_SPT3G}\texorpdfstring{SPT-3G$\times$ngWL}{SPT-3GxngWL} forecast}
In this section we present our SPT-3G$\times$ngWL forecasts for the four different models.
Figure~\ref{fig:posterior_fR0} shows the posterior distribution of the $\log_{10}|f_{R0}|$ parameter of the four models tested. The full posteriors of all cosmological parameters from the SPT-3G$\times$ngWL dataset are presented in Appendix~\ref{app:all_constraints}.
In fact, we have created three statistically independent mocks for each model to test the sensitivity of our conclusions to the statistical fluctuations that arise through the Poisson sampling of the HMF and sources of scatter in the observable--mass relations. As expected, no statistically significant difference is found between the three realizations per model, and we hence only show the posterior for one of the mock realizations here.

Table~\ref{tab:constraints_fR0} shows the constraints of the $\log_{10}|f_{R0}|$ parameter for the four analyzed models. 
The results of the $95\,\%$ upper bounds show very similar constraining power for the GR and F7 models and thus validate our choice of $\log_{10}|f_{R0}|=-7$ as a lower limit 
within the context of the SPT-3G dataset. Our analysis shows that an SPT-3G$\times$ngWL dataset can be used to distinguish GR and $f(R)$ modified gravity models down to $\log_{10}|f_{R0}| < -5.97$ ($95\,\%$ upper bound). 
\begin{figure}
        \centering
        \includegraphics[width=0.9\linewidth]{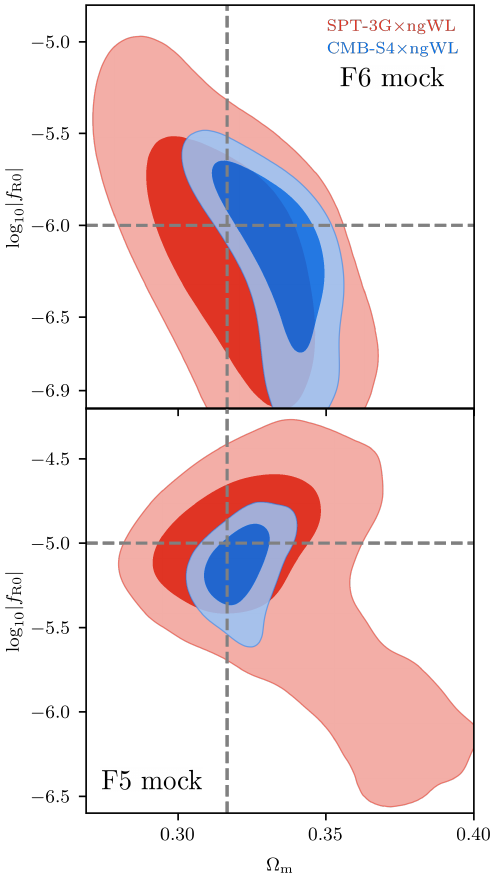}
        \vskip-0.25cm
        \caption{Constraints on $\log_{10}|f_{R0}|$ and $\Omega_{\mathrm{m}}$ for the F6 model (above) and F5 model (below) from the cluster abundance analysis for SPT-3G$\times$ngWL (red) and CMB-S4$\times$ngWL (blue) datasets. The parameter degeneracy that is clearly visible in the F6 panel arises due to enhanced cluster numbers in stronger $f(R)$ models being offset partially by lower $\Omega_\mathrm{m}$. In the F5 panel the degeneracy is still apparent at some level in the SPT-3G$\times$ngWL case and only mild but with orthogonal direction for the more constraining CMB-S4$\times$ngWL dataset. We tested the convergence of MCMC chain by running multiple chains with different settings and no shifts were seen.
        }
        \label{fig:panel_fR0_omegam}
\end{figure}

The $95\,\%$ credibility constraints from the F6 mock show that this cluster abundance and weak-lensing dataset without Planck\,2018 data can differentiate between GR and F6 models. Note, however, that models with a higher $\log_{10}|f_{R0}|$ value predict more clusters when leaving all other cosmological parameters unchanged (see Fig.~\ref{fig:mock_cat_dif_fR}) and these increased numbers translate into tighter constraints, as expected. 
Moreover, the SPT-3G$\times$ngWL dataset with Planck\,2018 data improves over the constraint from Ref.~\cite{Cataneo14} by $25\,\%$. 
Finally, the analysis of the F5 model shows that such modified gravity models could be distinguished from GR and from $f(R)$ gravity models with $\log_{10}|f_{R0}| < -5.97$ at $95\,\%$ credibility. 

As seen in Fig.~\ref{fig:posterior_fR0}, our analysis shows an asymmetric posterior distribution for $\log_{10}|f_{R0}|$ in the F5 model. An explanation for this asymmetric shape is given by the fact that the HMFs of weak modified gravity models are harder to distinguish because the differences are smaller. There is effectively less information on the $f(R)$ gravity parameter encoded in the HMF as one moves to weaker $f(R)$ models. Therefore, one expects a wider distribution or weaker constraints when going to smaller values of $\mathrm{log}_{10}|f_{R0}|$.

Figure~\ref{fig:panel_fR0_omegam} shows the degeneracy between $\log_{10}|f_{R0}|$ and $\Omega_{\mathrm{m}}$ for the F6 and F5 models. In the F6 model, the degeneracy is mild and vanishes when approaching smaller values of $\log_{10}|f_{R0}|$. The degeneracy between the two parameters is much stronger in the F5 model. 

\begin{figure}
        \centering
        \includegraphics[width=0.9\linewidth]{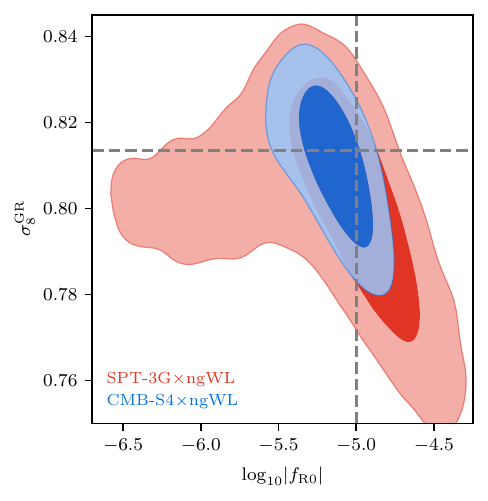}
        \vskip-0.25cm
        \caption{Joint constraints on $\log_{10}|f_{R0}|$ and 
        $\sigma_8^{\rm GR}$ of the F5 model from the cluster abundance analysis for SPT-3G$\times$ngWL (red) and CMB-S4$\times$ngWL (blue) datasets. The parameter degeneracy indicates that the more numerous clusters in stronger $f(R)$ models can be offset to some degree by a reduction in $\sigma_8^{\rm GR}$. This degeneracy is not seen in the F6, F7 and GR models for both datasets. We tested the convergence of MCMC chain by running multiple chains with different settings and no shifts were seen.}
        \label{fig:panel_fR0_sigma8}
\end{figure}

In addition, the analysis of the F5 model shows a degeneracy between $\log_{10}|f_{R0}|$ and $\sigma_8^{\rm GR}$ as seen in Fig.~\ref{fig:panel_fR0_sigma8}. This degeneracy vanishes for low $\log_{10}|f_{R0}|$ values for the SPT-3G$\times$ngWL dataset and the degeneracies are not seen in the analysis of the GR, F7 and F6 models.

The degeneracies between these parameters have been previously noted in, e.g., studies of $f(R)$ gravity using the weak-lensing power spectrum \cite{HarnoisDeraps22}.
These degeneracies arise from the fact that the three quantities $\log_{10}|f_{R0}|$ $\Omega_{\mathrm{m}}$ and $\sigma_8^{\rm GR}$ all change the amplitude of the HMF, enhancing the number of clusters for higher parameter values. The degeneracies are broken for small $\log_{10}|f_{R0}|$ values because the enhancement of the HMF in such $f(R)$ gravity models is very small to the point that the HMF is difficult to distinguish from GR. 

As stated in Sec.~\ref{subsec:HMF} there is a known degeneracy between the $f(R)$ gravity parameter and massive neutrinos \citep[see \eg][]{Motohashi13,Baldi14,Wright19}. 
Our analysis of the SPT-3G$\times$ngWL dataset on the other hand shows no degeneracy between these parameters for all four examined models. This is owing to the fact that in our analysis, the sum of neutrino masses is much smaller than the values considered in Refs.~\cite{Motohashi13,Baldi14,Wright19}, as we include the primary CMB constraint via our prior on $\Omega_\nu h^2$, and the impact of such low neutrino masses on the HMF is negligible. 

\subsubsection{\label{subsec:cosmo_CMBS4}\texorpdfstring{CMB-S4$\times$ngWL}{CMB-S4xngWL} forecast}
This section summarizes the results of our CMB-S4$\times$ngWL mock catalog analysis. As with the SPT-3G$\times$ngWL analysis, we analyze three statistically independent mocks and find no statistically significant differences. Figure~\ref{fig:posterior_fR0} shows the posterior distribution of the $\log_{10}|f_{R0}|$ parameter of the four models for one mock realization in blue. 
The full constraints in the cosmological parameters from the CMB-S4$\times$ngWL dataset are presented in Appendix~\ref{app:all_constraints}
 
The $\log_{10}|_{R0}|$ posteriors for the four analyzed models are shown in Table~\ref{tab:constraints_fR0}. One can see that the $\log_{10}|f_{R0}|$ upper limit for the GR model is slightly tighter (roughly $5\,\%$) than for the F7 model. This indicates that with the larger cluster sample of CMB-S4$\times$ngWL the few percent difference in the HMF as seen in Fig.~\ref{fig:HMF_fR} starts to play a role. 
In an actual analysis of the CMB-S4 cluster sample, one should thus extend the sampling range to lower values of $\log_{10}|f_{R0}|$.

Our results show that CMB-S4$\times$ngWL combined with Planck\,2018 data can rule out $f(R)$ gravity models down to $\log_{10}|f_{R0}| < -6.23$ at the $95\,\%$ upper limit. These models are currently still viable based on current published constraints. In fact, this forecast constraint represents an improvement of more than an order of magnitude compared to previously published cluster constraints~\cite{Cataneo14}.
Moreover, the analysis of the F6 model shows that a CMB-S4$\times$ngWL dataset can distinguish $\log_{10}|f_{R0}| = -6$ cosmology from a GR cosmology at the 
95\% credible interval. 
The tightest constraints from the CMB-S4$\times$ngWL mock dataset are for the F5 model because stronger modified gravity models predict more clusters.

\begingroup
\begin{table}
    \caption{
    Parameter posteriors for $\log_{10}|f_{R0}|$ in the four analyzed models in the two different clusters surveys SPT-3G$\times$ngWL and CMB-S4$\times$ngWL. For models F7 and GR we present $95\,\%$ upper limits whereas mean and $68\,\%$ credible intervals are shown for F6 and F5 models.}
    \begin{tabular}{lcc}
    \hline
    \hline
    
            Model \quad & \quad SPT-3G$\times$ngWL\quad  &\quad CMB-S4$\times$ngWL\quad  \\
    \hline
            GR  & $<-5.97$ & $<-6.23$  \\
            F7  & $<-5.98$ & $<-5.93$  \\
            F6  & $-6.12\pm 0.44$ & $-6.18\pm 0.31$  \\
            F5  & $-5.09\pm 0.43$ & $-5.13\pm0.17$  \\
    
    \hline
    \hline
    \label{tab:constraints_fR0}
    \end{tabular}
\end{table}
\endgroup

Figure~\ref{fig:panel_fR0_omegam} indicates a degeneracy between $\log_{10}|f_{R0}|$ and $\Omega_{\mathrm{m}}$ in the F6 and F5 models. 
In the case of the F5 model analysis, the degeneracy is mild and the direction is orthogonal to that from the analysis of the F6 model. 
The change in the degeneracy direction compared to the one in the F6 model can be explained by the behavior of the HMF at the high mass end for $\log_{10}|f_{R0}| = -5$ $f(R)$ gravity models. The change in the degeneracy is not seen in the F5 model for SPT-3G$\times$ngWL mock data because the contours are wider and leak into the low $\log_{10}|f_{R0}|$ region where the degeneracy changes direction. 
Furthermore the F5 model exhibits a degeneracy between $\log_{10}|f_{R0}|$ and $\sigma_8^{\rm GR}$ as seen in Fig.~\ref{fig:panel_fR0_sigma8}. As in the SPT-3G$\times$ngWL analysis, the F7 and GR models
do not show these two degeneracies, because the enhancement of the HMF vanishes for low values of $\log_{10}|f_{R0}|$ as discussed in Sec.~\ref{subsec:cosmo_SPT3G}.
In addition, the analysis of the CMB-S4$\times$ngWL datasets results in no significant degeneracy between $\log_{10}|f_{R0}|$ and $\Omega_\nu h^2$ for all models.

As we show in the last two sections, cluster data with weak-lensing informed mass calibration and with additional Planck\,2018 data in the weakest two $f(R)$ models will improve upon the best current constraints from cluster plus Planck data.

\section{\label{sec:discussion}Conclusions}
In this work, we present a forecast of cluster abundance constraints with weak-lensing informed mass calibration for a Hu \& Sawicki $f(R)$ model in two future surveys. 
In $f(R)$ gravity models, the clustering of matter is enhanced compared to the GR model due to a fifth force mediated by an extra scalar degree of freedom. Thus, the growth of structure is scale-dependent, and the collapse threshold for halos is correspondingly mass-dependent. We account for this in our analysis with a different HMF, which is given by the GR HMF scaled by an enhancement factor $\mathcal{R}$ involving the critical overdensity in $f(R)$ gravity $\delta_{\mathrm{crit}}$, and GR $\delta_{\mathrm{crit}, \mathrm{GR}}$ as well as its derivative with respect to $\mathrm{ln}M$ $\dd \delta_{\mathrm{crit}} / \dd \mathrm{ln} M$ calculated with a semi-analytical model.

The calculation of the critical overdensities and derivatives is computationally expensive and thus is not feasible for a cosmological analysis. Therefore, we build emulators for the three quantities needed to calculate the enhancement factor for the HMF. We show that the emulation error of the HMF is at the percent level and hence subdominant compared to other limitations. Using the emulators speeds up our calculation by more than three orders of magnitude. 

Using the $f(R)$ gravity HMF we analyze galaxy cluster datasets, which are a powerful cosmological probe able to distinguish between GR and $f(R)$ gravity due to the enhanced structure formation in $f(R)$ gravity. We create mock catalogs for the future tSZE cluster surveys representative of SPT-3G and CMB-S4, and we include weak-lensing mass calibration information from next-generation weak-lensing data from Euclid or Rubin. 
For each survey, we create three $f(R)$ gravity models with $\log_{10}|f_{R0}| \in [-7, -6, -5]$ (F7, F6 and F5 respectively) and a fourth GR model. We assume in our analysis that the F7 and GR models are indistinguishable because the HMF of the two models only differs by a few percent in the low mass regime, which is not easily distinguishable by either of these two future cluster surveys. In the case of the F7 and GR models, we add the Planck\,2018 CMB anisotropy data to improve the constraints. 

Our analysis consists of two steps. First, we carry out a weak-lensing mass calibration analysis of the observable--mass relations at a fixed, fiducial cosmology. Then we adopt the parameter posteriors from that analysis as priors on a cluster abundance analysis with the goal of deriving cosmological parameter posteriors.
We test the robustness of our pipeline by analyzing three statistically independent mock datasets for each of the models and find no statistically significant difference. The analysis of the four models of an SPT-3$\times$ngWL dataset gives the following results:
\begin{itemize}
    \item Cluster data from SPT-3G$\times$ngWL combined with primary CMB Planck\,2018 data improves the current best constraints from cluster data to $\log_{10}|f_{R0}| < -5.95$ at $95\,\%$ upper limit.
    \item The analysis of the F7 and GR model verifies that the two models are indistinguishable within the statistical uncertainties in the data.
    \item Our analysis shows that the F6 model is distinguishable from GR at the $95\,\%$ credible level.
    \item Modified gravity models with $\log_{10}|f_{R0}| = -5$ are distinguishable from GR and from $f(R)$ gravity models below $\log_{10}|f_{R0}| = -5.97$ at the $95\,\%$ credible level. 
    \item We observe degeneracies between $\log_{10}|f_{R0}|$ and $\Omega_{\mathrm{m}}$ as well as $\sigma_8^{\rm GR}$. These are broken for values of $|f_{R0}|$ below $10^{-6}$ and  $10^{-5}$ respectively.
\end{itemize}
The results of the CMB-S4$\times$ngWL mock data are the following: 
\begin{itemize}
    \item CMB-S4$\times$ngWL data plus Planck\,2018 data can rule out $f(R)$ gravity models above $\log_{10}|f_{R0}| = -6.23$ at $95\,\%$ upper bound. They not only improve upon the current best constraints from cluster data, but yield constraints on $|f_{R0}|$ that are lower by a factor of $0.38$ than the one from the SPT-3G$\times$ngWL dataset. 
    \item Modified gravity models with $\log_{10}|f_{R0}| = -6$ and $\log_{10}|f_{R0}| = -5$ are distinguishable from GR at the $95\,\%$ credible interval.
    \item The analysis of the F6 and F5 models show degeneracy between $\log_{10}|f_{R0}|$ and $\Omega_{\mathrm{m}}$ as well as $\sigma_8^{\mathrm{GR}}$.
    \item The degeneracy between $\log_{10}|f_{R0}|$ and $\Omega_{\mathrm{m}}$ in the analysis of the F5 model changes direction compared to the one in the F6 model.
    This is due to the different behavior of the HMF at the high mass end for the F5 model.
\end{itemize}

Overall our analysis shows that upcoming tSZE-selected cluster samples of thousands to tens of thousands of systems, combined with next-generation weak-lensing survey data such as that from Euclid or Rubin, will enable substantially improved constraints on modified gravity models. Furthermore, the analysis shows that constraints on $f(R)$ gravity from cluster and WL surveys are strongly competitive with other cosmological probes \cite{Cataneo14,Hojjati16,Hu16,Pratten16,Yamamoto10,Stark16}.

\begin{acknowledgments}
We thank Baojiu Li, César Hernández-Aguayo and Cheng-Zong Ruan for sharing their FORGE HMF emulator with us.
This research was supported by 1) the Excellence Cluster ORIGINS, which is funded by the Deutsche Forschungsgemeinschaft (DFG, German Research Foundation) under Germany's Excellence Strategy - EXC-2094-390783311, 
by 2) the Max Planck Society Faculty Fellowship program at MPE, and by 3) the Ludwig-Maximilians-Universit\"at in Munich.

\end{acknowledgments}

\appendix
\section{\label{app:better_WL_uncertainties}The impact of improved WL systematic uncertainties on the mass calibration} 
\begin{figure*}
    \centering
    \includegraphics[width=\textwidth]{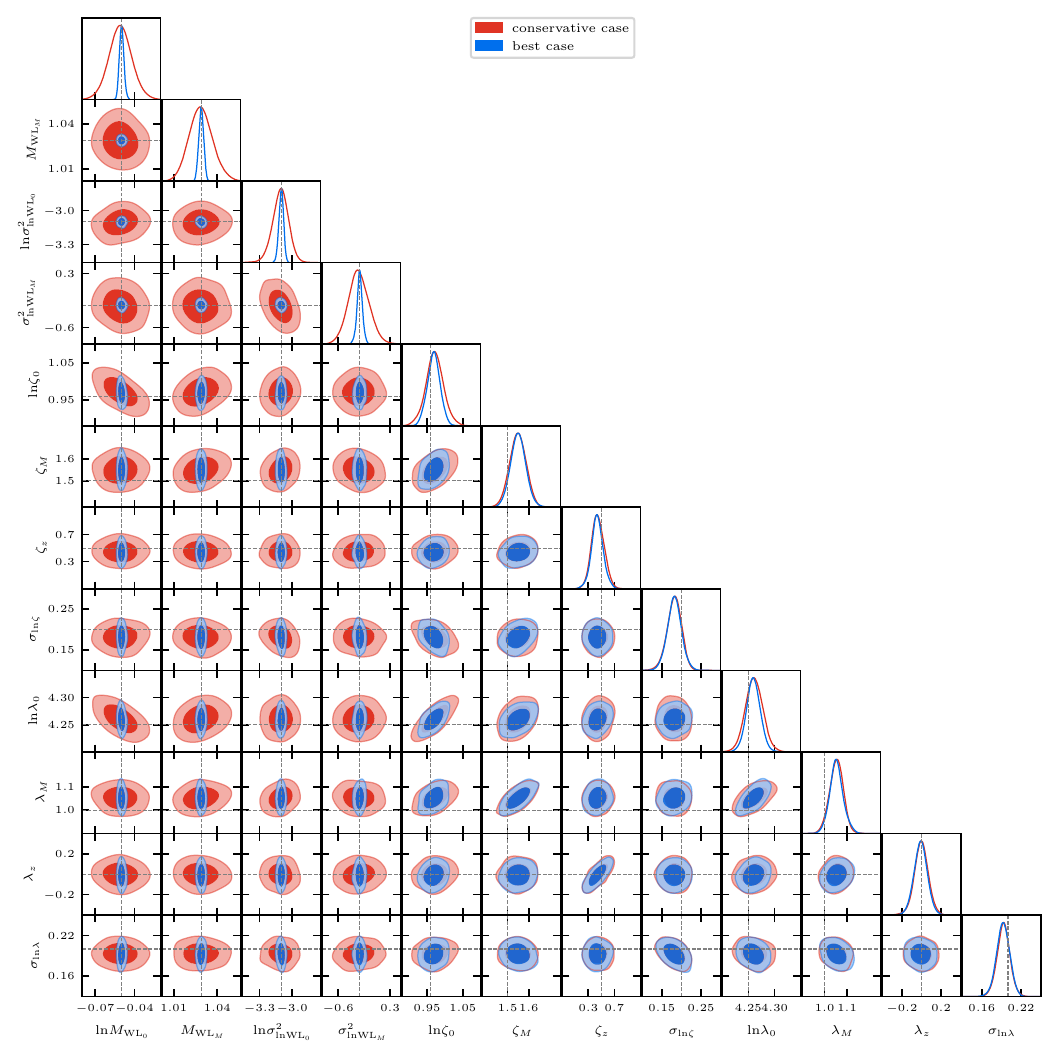}
    \caption{Mass calibration analysis with two different choices on the weak-lensing uncertainties (WL systematic floor) for the ngWL survey. The conservative choice in red has a factor of two improvement in WL systematics compared to DES, while the optimistic case is a factor of ten better than DES. The posteriors on the observable mass scaling relation parameters are hardly impacted, indicating that in neither case are the posteriors dominated by the systematic uncertainties in the weak-lensing analysis.}
    \label{fig:Euclid_WL_uncert_compare}
\end{figure*}
\begin{figure*}
    \centering
    \includegraphics[width=\textwidth]{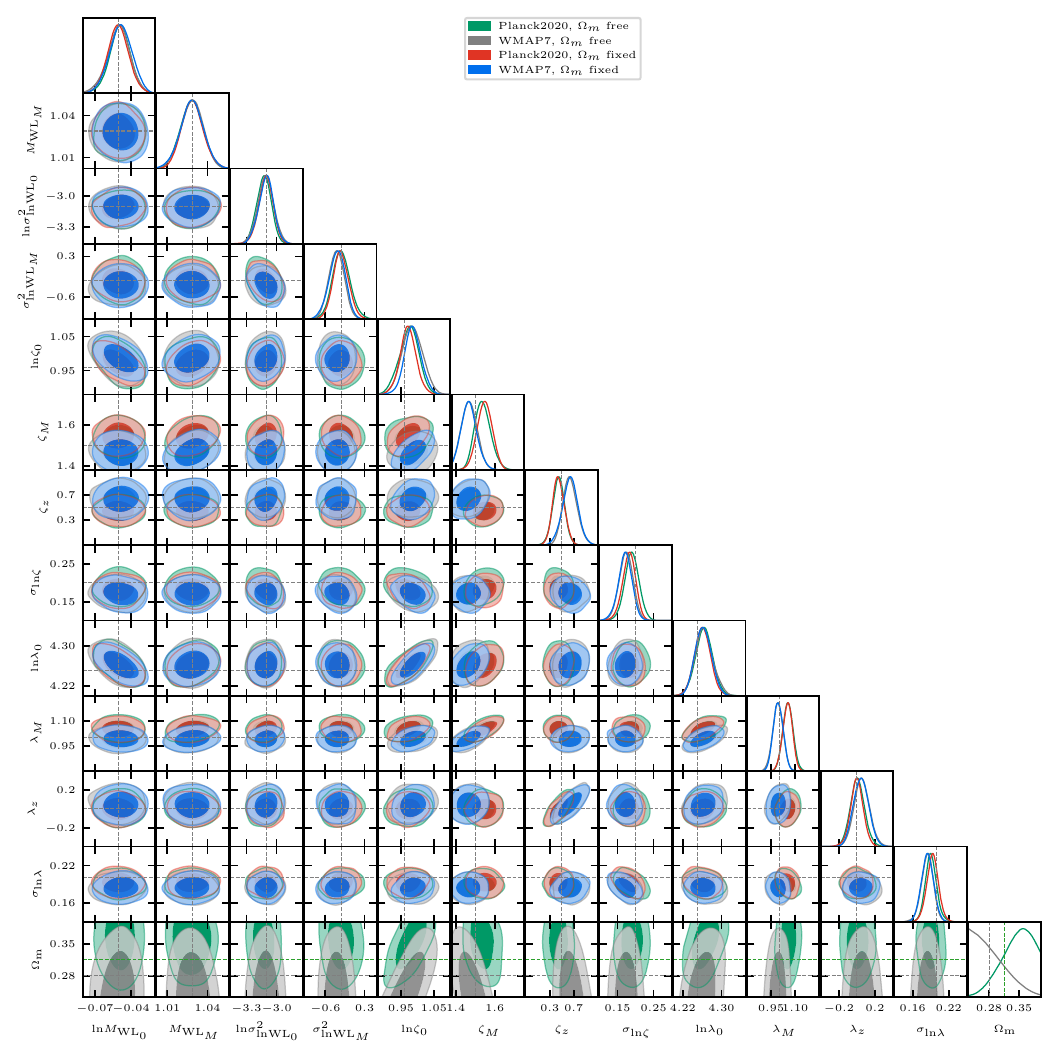}
   
    \caption{Mass calibration analysis with two different cosmologies: Planck\,2018 (green and red) and WMAP7 (gray and blue). For each cosmology, we execute a chain with $\Omega_{\mathrm{m}}$ fixed (red and blue) and free (green and gray). Dashed lines indicate input values, for $\Omega_{\mathrm{m}}$ the input value is represented in green for the Planck\,2018 mock and in gray for the WMAP7 mock. The consistency of the observable--mass scaling relation parameter posteriors indicates that neither the exact input value of $\Omega_\mathrm{m}$ used in the mocks nor fixing it or fitting $\Omega_\mathrm{m}$ have an important impact.}
    \label{fig:mass_calib_diff_cosmo}
\end{figure*}

In this appendix, we investigate the dependence of our results on the assumptions made about systematics in the weak lensing mass estimates.
Next-generation weak-lensing surveys as expected from the Euclid satellite and the Vera Rubin Observatory will collect a much higher amount of data compared to current surveys like DES, KiDS and HSC SSP \cite{Aihara18}. Therefore, much work will be done to improve the systematic uncertainties to take advantage of the increasing amount of data for the mass calibration analysis.

For photometric redshift and shear systematics, we simply adopt the requirements for the ngWL surveys, which are dramatically tighter than what has been achieved in the ongoing surveys. But, as discussed in Section~\ref{sec:mocks}, there are also systematics associated with how we use the observed source galaxy shear profiles to determine the weak-lensing mass $M_\mathrm{WL}$, and we model these in the so-called weak-lensing to halo mass relation (Eqs.~\eqref{eq:WL_mass_rel} and \eqref{eq:WL_mass_var}). The exact uncertainties in the parameters of this relation are estimated for the ngWL surveys based on the current DES results \cite{Bocquet23}.
    \begin{figure*}
        \centering
        \includegraphics[width=\textwidth]{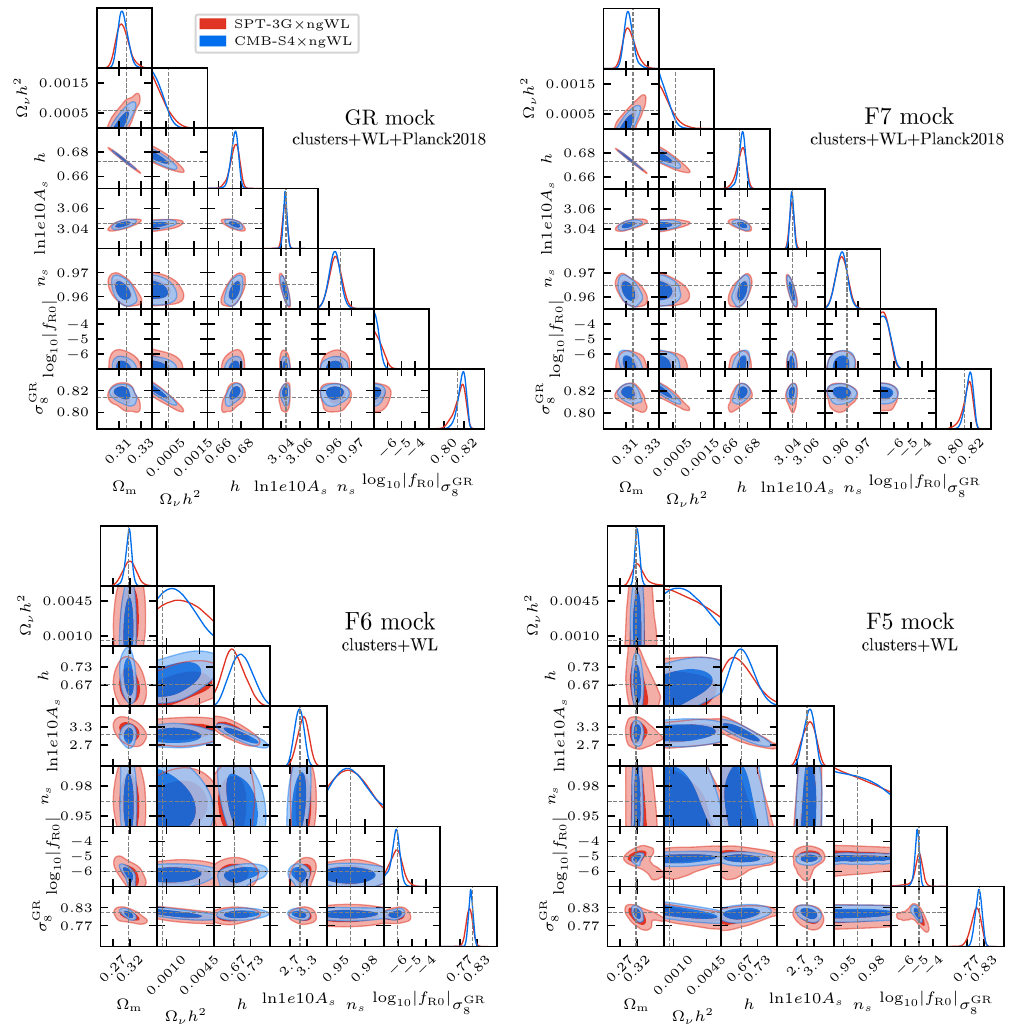}
        \caption{Posterior distributions of the cluster abundance analysis for SPT-3G$\times$ngWL (red) and CMB-S4$\times$ngWL (blue) datasets. For the analyses of the GR and F7 models, the Planck\,2018 anisotropy data were included.}
        \label{fig:triangles_both_surveys}
    \end{figure*}

We assume a baseline (conservative) approach where the uncertainties on the parameters in Eqs.~\eqref{eq:WL_mass_rel} and \eqref{eq:WL_mass_var} would be improved by a factor of two as compared to DES, and an optimistic scenario where the uncertainties would be reduced by a factor of ten compared to DES. 

The comparison of the conservative baseline and our optimistic scenario is shown in Fig.~\ref{fig:Euclid_WL_uncert_compare}. One can see that the reduction of the WL uncertainties (corresponding to a kind of systematic floor) by a factor of five between our conservative and optimistic scenarios does not significantly impact the scaling relation parameter posteriors.
We do see an improvement of around $5\,\%$ in the constraints on the parameters $\ln\asz$ and $\ln\alambda$, which indicates that the systematic floor in our conservative baseline is having some impact. Qualitatively, we expect this by looking at the contours of Fig.~\ref{fig:Euclid_WL_uncert_compare} between the scaling relation parameters and the WL uncertainties. Besides the contours of $\bWL$ and $\ln\asz$ or $\ln \alambda$ respectively, the contours are horizontally orientated. This is an indication that an improvement of the WL uncertainties, which only shrinks the contours in the horizontal directions, will not lead to better constraints on the scaling parameters. In the case of the two amplitude parameters, the contour plot shows a small degeneracy between $\bWL$ and $\ln\asz$ or $\ln \alambda$, respectively, which helps explain the small improvement of these two amplitude parameters when reducing the systematic floor by a factor of five.

\section{\label{app:cosmos_inteo_mass_calibration}Cosmology-dependence of mass calibration}
As mentioned in Section~\ref{subsec:mass_calibration_constraints} we assume that the ngWL based mass calibration analysis can be separated from the SPT-3G and CMB-S4 abundance analysis without introducing any important biases. This assumption is driven by our need to avoid running the mass calibration, which is the bottleneck in the current version of our code, many different times. We verify this statement by running the mass calibration analysis with both fixed and free matter density parameter $\Omega_{\mathrm{m}}$ with our standard Planck\,2018 priors. Freedom in $\Omega_{\mathrm{m}}$ has the most important cosmological impact on our mass calibration analysis because the mapping from shear profiles to halo mass depends on the distance-redshift relation. In addition, we analyze a GR mock with a WMAP7 cosmology \cite{Komatsu2011}, \ie $\Omega_{\mathrm{m}}= 0.27$, $\Omega_{\mathrm{b}}= 0.0469$, $h=0.7$, $\mathrm{ln}(10^{10}A_s) = 3.155$ and $n_s = 0.95$ with $\Omega_{\mathrm{m}}$ fixed or free. This tests whether there is sensitivity to the precise input value of the cosmological parameters. 

The results for the mass calibration for all four runs give the same constraints on the observable--mass scaling relation parameters. Figure~\ref{fig:mass_calib_diff_cosmo} includes the posteriors from these four runs (color coded as marked in the figure). There is no tension among the four sets of posteriors, aside from the $\Omega_\mathrm{m}$ posterior where the input Planck\,2018 and WMAP7 values are indeed different. This behavior motivates our assumption that the mass calibration is independent of the cosmology and therefore validates our approach of separating the mass calibration and cosmology chains in the analyses we present here. 

\section{\label{app:all_constraints}Full results for the \texorpdfstring{SPT-3G$\times$ngWL}{SPT-3GxngWL} and \texorpdfstring{CMB-S4$\times$ngWL}{CMB-S4xngWL} datasets}
In Fig.~\ref{fig:triangles_both_surveys} we show the constraints on all cosmological parameters for all four examined models for the SPT-3G$\times$ngWL and CMB-S4$\times$ngWL datasets in red and blue, respectively. 

\bibliography{apssamp}

\end{document}